\documentclass[final,3p,times]{elsarticle}
\usepackage{amssymb}
\usepackage{amsmath}
\usepackage{bm}
\usepackage{amsthm}
\usepackage[ruled,vlined]{algorithm2e}
\SetKwProg{Init}{Initialization:}{}{}
\usepackage{caption}
\usepackage{subcaption}
\usepackage{algpseudocode}
\usepackage{tabularx}
\usepackage{float}
\usepackage{comment}
\usepackage{siunitx}
\usepackage{hyperref}
\usepackage[utf8]{inputenc}
\usepackage{xcolor}
\newcommand{\todo}[1]{\textcolor{red}{TODO:#1}}

\usepackage{url}
\usepackage{paralist}
\begin{document}

\begin{frontmatter}



\title{Parallel Computation of Inverse Compton Scattering Radiation Spectra based on Liénard-Wiechert Potentials}


\author[desy,uhh]{Yi-Kai Kan}
\author[desy,uhh]{Franz X. Kärtner}
\author[tuhh]{Sabine Le Borne}
\author[tuhh]{Daniel Ruprecht}
\author[tuhh]{Jens-Peter M. Zemke}

\affiliation[desy]{organization={Center for Free-Electron Laser Science, Deutsches Elektronen-Synchrotron DESY},%Department and Organization
            addressline={Notkestraße 85}, 
            city={22607 Hamburg},
            country={Germany}}
            
\affiliation[uhh]{organization={Department of Physics, University of Hamburg},%Department and Organization
            addressline={Luruper Chaussee 149},  
            city={22761 Hamburg},
            country={Germany}}          
            
\affiliation[tuhh]{organization={Hamburg University of Technology, Institute of Mathematics},%Department and Organization
            addressline={Am Schwarzenberg-Campus 3}, 
            city={21073 Hamburg},
            country={Germany}}

\begin{abstract}
Inverse Compton Scattering (ICS) has gained much attention recently because of its promise for the development of table-top-size X-ray light sources. 
Precise and fast simulation is an indispensable tool for predicting the radiation property of a given machine design and to optimize its parameters. 
Instead of the conventional approach to compute radiation spectra which directly evaluates the discretized Fourier integral of the Liénard-Wiechert field given analytically (referred to as the frequency-domain method), this article focuses on an approach where the field is recorded along the observer time on a uniform time grid which is then used to compute the radiation spectra after completion of the simulation, referred to as the time-domain method. 
Besides the derivation and implementation details of the proposed method, we analyze possible parallelization schemes and compare the parallel performance of the proposed time-domain method with the frequency-domain method. We will characterize scenarios/conditions under which one method is expected to outperform the other. 
\end{abstract}


%%Research highlightshttps://www.overleaf.com/project/60a4b400dbf03159b50cb264
% \begin{highlights}
% \item Numerical simulation of inverse Compton scattering for the development of table-top size X-ray light sources.
% \item An efficient discretization and implementation of the time-domain method for solving Liénard-Wiechert potentials.
% %\done{JZ: I think we (essentially you, as I will most probably not be there on Thursday) need to figure out our “research highlight(s)”, or?}
% \item A parallelization scheme of the time-domain method for solving Liénard-Wiechert potentials.
% \item Numerical performance models and comparison of the time- and frequency-domain methods for solving Liénard-Wiechert potentials for the inverse Compton scattering process.
% \end{highlights}

\begin{keyword}
Liénard-Wiechert Field \sep Radiation Spectra \sep Inverse Compton Scattering
\PACS 41.60.-m  % Radiation by moving charges
\sep 02.70.-c   % Computational techniques simulations 
\MSC 78A40 %Waves and radiation in optics and electromagnetic theory
\sep 65Y05 %Parallel numerical computation
\sep 65Y20 %Complexity and performance of numerical algorithms
\sep 65D05 %Numerical interpolation
\sep 42A15 %Trigonometric interpolation

\end{keyword}

\end{frontmatter}


\section{Introduction}
The most powerful X-ray sources today are Free-Electron Lasers (FEL) built at large national facilities producing intense, spatially coherent X-ray radiation \cite{bostedt_linac_2016,pellegrini_development_2020}.
FELs rely on magnetic undulators with periods on the order of 10 cm and therefore need highly relativistic beams to reach the hard X-ray regime. Inverse Compton scattering (ICS) instead uses an optical undulator, i.\,e., a counter propagating laser pulse, and therefore needs only weakly relativistic beams for hard X-ray production. ICS sources have therefore drawn great attention over the last ten years because of the potential to allow for table-top-size hard X-ray light sources with a much improved performance~\cite{krafft_compton_2010, ta_phuoc_all-optical_2012} when compared to an X-ray tube, eventually approaching that of a second generation synchrotron \cite{bilderback_review_2005}. 
Such X-ray light sources would greatly increase access to high brightness X-ray radiation for fast experimental turn-around rather than waiting many months for using the light-source provided by large national facilities~ \cite{graves_compact_2014} including international travel to these facilities. Structural biology, biomedical imaging and materials screening would greatly benefit from such developments \cite{kartner_axsis_2016}.

\begin{figure}[th]
    \centering
    \includegraphics[width=0.5\textwidth]{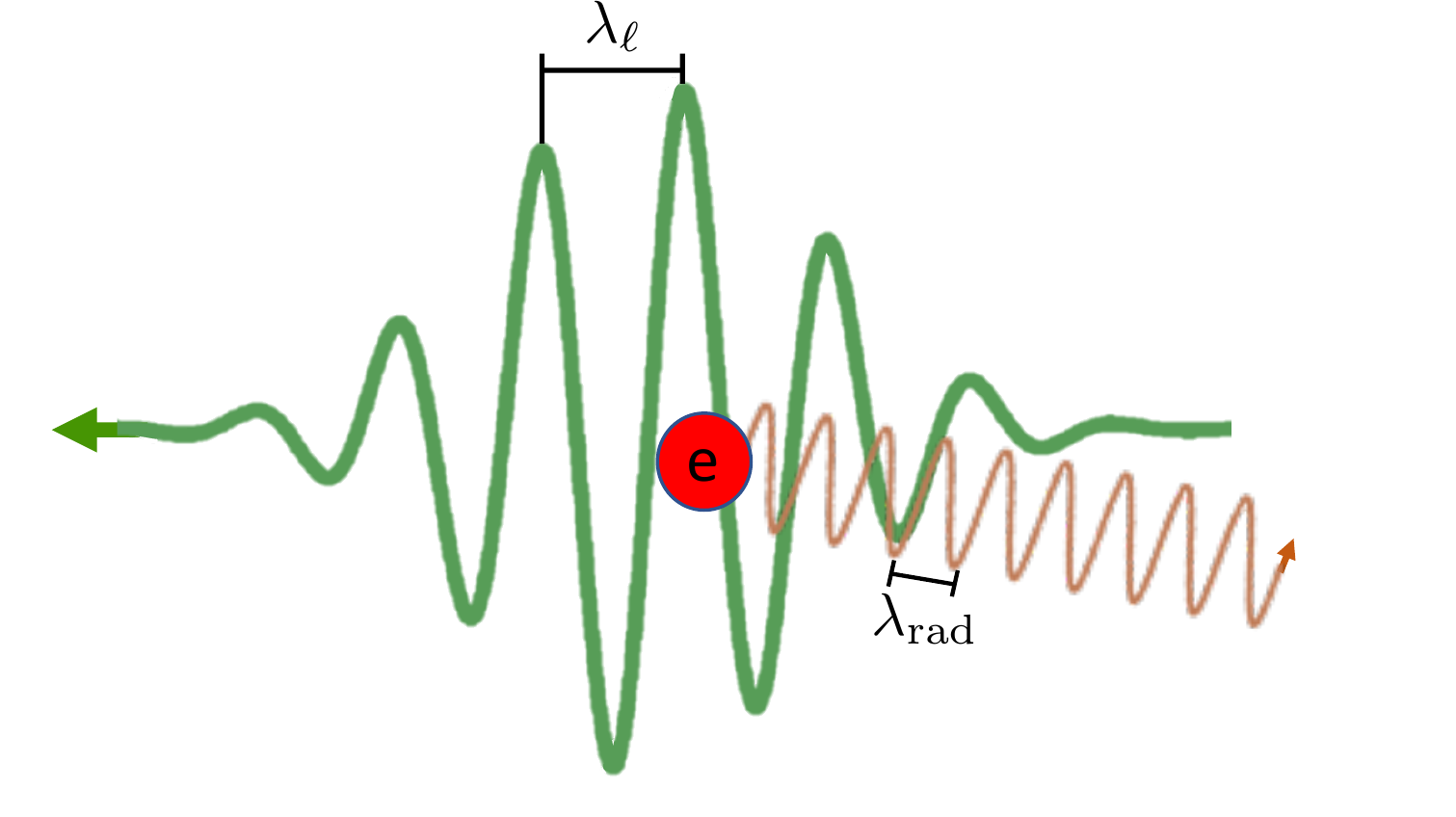}
    \caption{Illustration of the ICS process. An electron (red circle) collides with a counter-propagating laser pulse (green line) of wavelength $\lambda_{\ell}$ and generates radiation (brown line) with shorter wavelength $\lambda_{\text{rad}}=(1+a^2_0/2)\lambda_{\ell}/4\gamma^2$.}
    \label{fig:ics_schematic}
\end{figure}

In ICS, an electron bunch collides purposefully with a counter-propagating high-intensity laser pulse. 
The electrons are driven by the oscillating electrical field of the laser pulse to undergo a wiggling motion (Fig.~\ref{fig:ics_schematic}). 
The radiation wavelength with maximum energy emitted from wiggling electrons is given by
\begin{equation*}
  \lambda_{\text{rad}} = \dfrac{\lambda_{\ell}}{4\gamma^2}\left(1+\dfrac{a^2_{0}}{2}\right)
\end{equation*}
where $\gamma$ is the electron energy (normalized to its energy at rest), $\lambda_{\ell}$ is the laser wavelength and $a_{0}$ is the normalized vector potential characterizing the strength of the laser field~\cite{esarey_nonlinear_1993}. 
As already discussed above, the counter-propagating laser pulse is often called an optical undulator with an equivalent period much shorter than that of magnetic undulators in conventional FELs. 
Thus, being able to achieve short wavelength radiation without a high-energy electron bunch driven by a large-scale accelerator is the major advantage of ICS sources. 

Due to the increasing demand on developing ICS sources~\cite{kartner_axsis_2016, deitrick_high_2017, deitrick_high-brilliance_2018-1, graves_asu_2018, graves_compact_2014, brummer_design_2020, gunther_versatile_2020, variola_thomx_nodate, dupraz_thomx_2020}, numerical simulation is an indispensable tool to understand the generated radiation characteristics. 
To simulate the generated radiation, the Liénard-Wiechert field method is among the commonly used methods~\cite{thomas_algorithm_2010, frederiksen_radiation_2010, haugbolle_photon-plasma_2013, pausch_computing_2014, pausch_how_2014}. 
In this method, the radiation field is computed from the charged particle trajectories which are either given beforehand or solved in parallel during the simulation. 

One of the challenges when computing radiation spectra generated by relativistic particles lies in the delay between the particle generating its contribution (retarded or emission time) and the observer detecting it (advanced time) which is not constant but depends on the distance between the particle and the observer at time of emission.
Furthermore, to allow for a simple and quick computation of spectra via fast Fourier transform, it is preferable to have contributions at the observer on a uniform, equidistant time mesh.
One possibility is to define a uniform mesh in the advanced time and then calculate the corresponding emission times (“retarded time scheme”).
However, this approach is computationally expensive as it requires (numerical) root-finding and it is also memory intensive as full particle trajectories must be stored to allow interpolation to the computed emission times.
By contrast, the advanced time scheme computes particle trajectories and emissions using a uniform time grid.
At each time step, the corresponding advanced time is computed when the generated emission reaches the observer.
This approach avoids computationally expensive root-finding but raises the new problem of how to deal with the different (non-uniform) advanced times at which the emissions reach the observer.
Since one is typically interested in the amplitude $\left| E(x,\omega) \right|$ of the generated field in the frequency-domain~\cite{pausch_electromagnetic_2012} and not $E(x,t)$ itself, instead of first depositing contributions in physical space on a mesh and then applying a Fourier transform, it is possible to directly compute the amplitude that the particle contributes to each Fourier mode. This approach is commonly referred to as the frequency-domain method since it calculates directly the spectrum of the generated field but never the field itself~\cite{thomas_algorithm_2010,pausch_electromagnetic_2012}.
However, as the computation of each Fourier mode is performed independently, this method intrinsically needs a high number of operations, i.\,e., $\text{O}(N_{\omega}\cdot N_{p}\cdot N_{s}$) where $N_{\omega}$, $N_{p}$ and $N_{s}$ are the number of frequency points for the spectrum, the number of simulated particles and the number of simulation steps, respectively \cite{pausch_electromagnetic_2012}.
As an alternative, Sell et al.~\cite{sell_attosecond_2014} use a time-domain approach where particle contributions at a given advanced time are interpolated to a uniform observer time mesh before being (Fourier) transformed to the frequency domain.

 This method can benefit from the fast computation of the overall discrete spectrum using the Fast Fourier Transform (FFT) because the full sequence of interpolated fields over the observer time mesh is known. Thus, the time-domain method may have advantages in terms of computational time. This was pointed out in \cite{pausch_electromagnetic_2012}, however without a detailed analysis and verification.

In spite of the high computational cost, the frequency-domain method can be efficient in terms of memory cost when a large number of particles needs to be considered~\cite{pausch_electromagnetic_2012}. In this scenario, it may not be possible nor necessary to store the full history of all particle trajectories and the calculation of radiation spectra can be done in parallel to the calculation of particle trajectories in a simulation. 
To study real-world cases where a large number of particles and observation points are usually needed, the use of high-performance computers may be necessary~ \cite{pausch_computing_2014, pausch_how_2014}.

For the time-domain method, the superposition of particle fields relies on the interpolation onto a predefined uniform (advanced) time mesh. This implies that the particle information in some previous steps is needed when determining the superimposed field. In the implementation proposed by~\cite{pausch_electromagnetic_2012}, the full history of the particle trajectories is available beforehand. Under this assumption, the computation of the particle trajectories is completely detached from the computation of the radiation field. However, for a simulation with a large number of particles, the storage of their trajectories needs an excessive memory capacity and such simulation may not be possible in practice. To mitigate the memory consumption for storing the particle trajectories, an algorithm which solves the particle trajectory and computes the superimposed field simultaneously might be necessary. One approach belonging to this category can be found in~\cite{sell_attosecond_2014}. 
%Their approach has been parallelized within a shared memory architecture, and the memory limitation %is a constraint on designing the algorithm. 
%Instead of storing each particle's arrival field and the arrival time at previous time step for each %observer, 
The particle trajectories are stored only for a certain number of preceding time steps in a so-called “ring buffer” \cite{sell_attosecond_2014} and used to interpolate the field onto a pre-defined uniform time grid. The drawback of this approach is that typically more field evaluations are performed (and stored) than are necessary for the interpolation.

In this article, we first describe the frequency-domain method to compute radiation spectra based on the Liénard-Wiechert potential. 
After that, we introduce our discretization and implementation of the time-domain approach and provide an analysis for two possible distributed parallelization schemes.
In particular, the proposed algorithm needs to store the particle trajectories only for a single preceding time step. 
Finally, we propose and discuss a strategy for choosing parameters when applying the time-domain method and the frequency-domain method to compute ICS radiation spectra. Following this strategy, we analyze the performance of both methods and discuss the scenarios where one method outperforms the other. We conclude that the time-domain method is in general, i.\,e., within the specification of real-world  experiment projects, more favorable than the frequency-domain method in terms of execution time in serial and in parallel when applied to compute the radiation spectra of an ICS process. Besides, we also show that the frequency-domain method can outperform the time-domain method in some circumstances.

\section{Radiation Calculation using Liénard-Wiechert Fields}
\begin{figure}[th]
\centering
\includegraphics[width=0.30\textwidth]{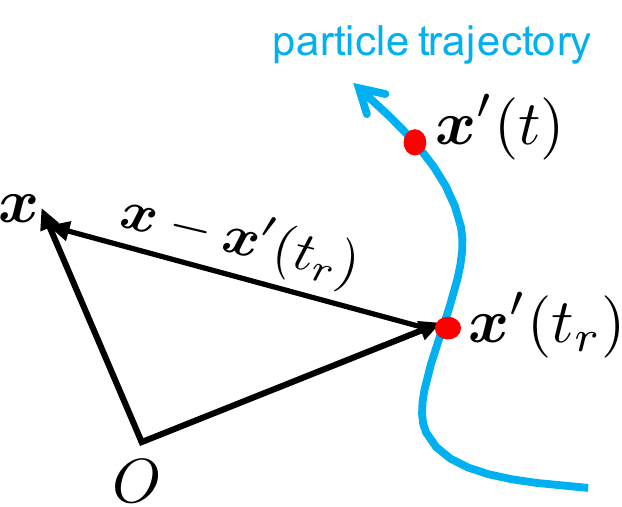}
\caption{Illustration of the retardation condition~\eqref{eq:retarted_condition}. Radiation emitted from a particle at time $t_r$ at position $\bm{x}'(t_r)$ will arrive at the observer at position $\bm{x}$ at time $t$. The travel time of the radiation is $|\bm{x}-\bm{x}'(t_r)|/c_0$.}
\label{fig:schematic_lienard_wiechert}
\end{figure}

The radiation field that is emitted from a charged particle at position $\bm{x}'(t_{r})$ at the (retarded) time $t_{r}$ and observed at a fixed position $\bm{x}$ at time $t$ (see Fig.~\ref{fig:schematic_lienard_wiechert}) is given by the equations
\begin{equation}
  \label{eq:lw_fields_retarded}
  \bm{E}(\bm{x},t)
  = \frac{1}{4\pi\varepsilon_{0}}\cdot
    \frac{q\bm{n}\times\left((\bm{n}-\bm{\beta}'(t_r))
    \times\bm{\dot{\beta}'(t_r)}\right)}
    {c_{0}(1-\bm{n}\cdot\bm{\beta}'(t_r))^3\left|\bm{x}-\bm{x}'(t_r)\right|}, \qquad
    \bm{B}(\bm{x},t)
  = \frac{\bm{n}}{c_{0}}\times\bm{E}(\bm{x},t),
\end{equation}
%where $\bm{x}$ is position of observation, and $\bm{x}'$ is the particle position.
where $\bm{\beta}'\equiv\bm{x}'/c_0$ and $\dot{\bm{\beta}}'\equiv\ddot{\bm{x}}'/c_0$ denote the particle's velocity and acceleration (normalized by the speed of light $c_0$ in vacuum), $q$ is the charge of the particle, $\varepsilon_{0}$ is the vacuum permittivity, defined by $\varepsilon_{0}\equiv1/(\mu_0c_0^2)$ for the vacuum permeability $\mu_0$, and $\bm{n}\equiv(\bm{x}-\bm{x}')/|\bm{x}-\bm{x}'|$. (Compared to ~\cite[Eqs.~(14.13)+(14.14)] {jackson_classical_nodate}, the velocity field (14.14) can be neglected in the radiation problem since the total power of the velocity field decays with the distance.)
%\done{JZ: In Jackson's textbook an additional term occurs, which is neglected here due to the factor $R/\gamma$, where $R$ denotes the radius (observers on a sphere). Should we remark on that? This is related to the applicability of our methods. The first term and second term in eqn (14.14) are velocity and acceleration field respectively. The acceleration field is usually referred to radiation field because it's magnitude falls off ~1/R. This means the radiation power $|E|^2\cdot 4\pi R^2$ (integrate over a sphere surface) is independent of $R$ and this mean only the energy.}
The observation time $t$ and (retarded) emission time $t_r$ fulfill the retardation condition (see Fig.~\ref{fig:schematic_lienard_wiechert})
\begin{equation}
    t = t_{r} + \dfrac{|\bm{x}-\bm{x}'(t_r)|}{c_{0}},
    \label{eq:retarted_condition}
\end{equation}
{i.e.}, the radiation field generated at position $x'(t_r)$ at time $t_{r}$ travels with the speed of light $c_0$ to reach the observation point $\bm{x}$.
%. It takes $\tfrac{|\bm{x}-\bm{x}'(t_r)|}{c_{0}}$ for the radiation field to travel to the observation point where the signal will arrive 
at $t = t_{r} + |\bm{x}-\bm{x}'(t_r)|/c_{0}$.
There are some drawbacks when applying Eqns.~\eqref{eq:lw_fields_retarded},~\eqref{eq:retarted_condition} to compute the radiation field:
\begin{compactitem}
\item Root finding is needed to solve the retardation condition~\eqref{eq:retarted_condition} for $t_r$ which is computationally intensive~\cite{ryne_using_2013}.
\item The trajectories of all particles have to be stored since they are required to compute the radiation field.
\item In a numerical simulation, the electron trajectories are computed at discrete time points and interpolation is needed when $t_r$ falls between two consecutive time points~\cite{ryne_using_2013}.
\end{compactitem}

\begin{figure}[th]
\centering
\begin{subfigure}[b]{0.45\textwidth}
    \includegraphics[width=\linewidth]{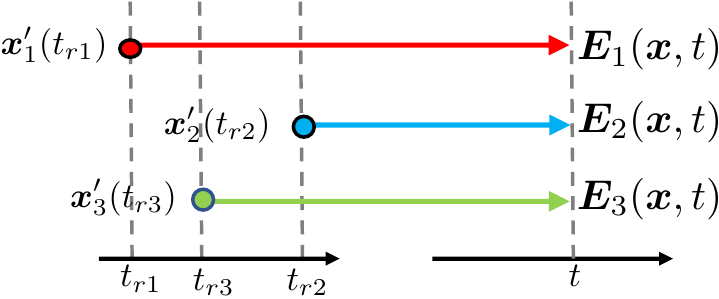}
    \caption{retarted time scheme}
\end{subfigure}
\hspace*{1em}
\begin{subfigure}[b]{0.45\textwidth}
    \includegraphics[width=\linewidth]{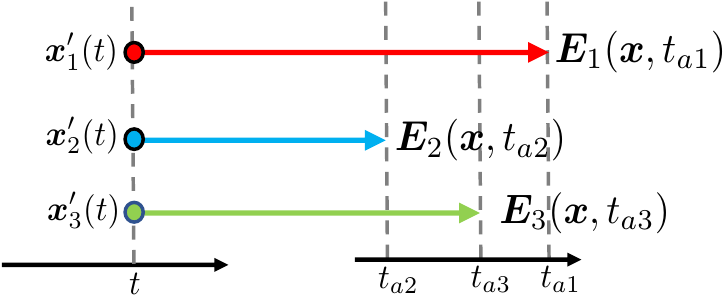}
    \caption{advanced time scheme}
\end{subfigure}
\caption{Illustration of (a) retarded time scheme and (b) advanced time scheme. Emission of radiation from three particles is considered and represented by different colors. To compute the field arriving at the observer $\bm{x}$ at time $t$ from each particle in the retarded time scheme, we need to solve Eq.~\eqref{eq:retarted_condition} for the retarded times $t_{r1}$, $t_{r2}$ and $t_{r3}$. In the advanced time scheme, the particle trajectory at time $t$ is used to compute the fields at advanced times $t_{a1}$, $t_{a2}$ and $t_{a3}$, obtained by evaluating Eq~\eqref{eq:advanced_condition} at which the observer at $\bm{x}$ receives the radiation field.}
\label{fig:schematic_retarded_advanced}
\end{figure}

Alternatively, one may evaluate Eq.~\eqref{eq:lw_fields_retarded} at a future (advanced) time $t_a$, 
\begin{equation}
    t_a = t + \frac{|\bm{x}-\bm{x}'(t)|}{c_{0}},
    \label{eq:advanced_condition}
\end{equation}
to obtain the field generated by an electron's motion at the current time $t$, using the 
substitutions $t\to t_a$ and $t_r\to t$. 
This scheme does not involve root-finding and we can compute the time $t^{k}_{a}$ at which the radiation from particle $k$ arrives at the observation position $\bm{x}$.
However, as implied by Eq.~\eqref{eq:advanced_condition}, the arrival time of the radiation can be different for different particles and the superposition of particle fields is not straightforward.
Both schemes are illustrated in Fig.~\ref{fig:schematic_retarded_advanced}.

When studying radiation phenomena, it is usually valid to consider the far-field approximation, that is the radiation field from a particle which is observed at a distance far from its position $|\bm{x}|\gg|\bm{x}'(t)|$, we have
$$
  |\bm{x}-\bm{x}'(t)|\approx|\bm{x}|-\bm{n}\cdot\bm{x}'(t)
$$
where $\bm{n}\equiv\bm{x}/|\bm{x}|$, by abuse of notation, denotes an approximation of the normal vector $\bm{n}=(\bm{x}-\bm{x}'(t))/|\bm{x}-\bm{x}'(t)|$. 
The electric field and corresponding advanced time condition can be approximated accordingly,
\begin{equation}
  \bm{E}(\bm{x},t_a) 
  \approx\frac{1}{4\pi\varepsilon_{0}}\cdot
  \frac{q\bm{n}\times
  \left((\bm{n}-\bm{\beta}'(t))
  \times\bm{\dot{\beta}'(t)}\right)}{c_{0}
  (1-\bm{n}\cdot\bm{\beta}'(t))^3|\bm{x}|}\label{eq:far_efield},
\end{equation}
for
\begin{equation}
  t_a \approx t +\frac{|\bm{x}|}{c_0}
  -\frac{\bm{n}\cdot\bm{x}'(t)}{c_0}.\label{eq:far_adv}
\end{equation}
This approximation will be used throughout this article.

\section{Frequency-Domain Method}
One way to avoid the root-finding problem is to superimpose the fields at the observer in the frequency-domain. 
In a typical radiation study, one is interested in the spectral-angular distribution of the radiation~\cite[Eqn.~(14.60)]{jackson_classical_nodate}
\begin{equation*}
  \frac{dI^2}{d\omega d\Omega}=\frac{2}{c_0\mu_0}|\bm{x}|^2|\bm{E}(\bm{x},\omega)|^2,
\end{equation*}
where $\omega$ is the frequency and $\Omega$ is the solid angle of the observation sphere surface with radius $|\bm{x}|$. 
The electric field $E(\bm{x},\omega)$ in the frequency-domain is
\begin{equation}
    \bm{E}(\bm{x},\omega)
    \equiv\frac{1}{\sqrt{2\pi}}
    \int^{\infty}_{-\infty}\bm{E}(\bm{x},t)\exp(j\omega t)\,dt
    \label{eq:efield_xfrom}
\end{equation} 
where $j\equiv\sqrt{-1}$. 

By the superposition principle, the spectral-angular radiation distribution from a bunch of electrons is given by
\begin{equation}
  \label{eq:far_bunch_inty}
  \frac{dI^2}{d\omega d\Omega}
  = \frac{2}{c_0\mu_0}|\bm{x}|^2
  \Biggl|\sum^{N_p}_{k=1} \bm{E}_{k}(\bm{x},\omega)\Biggr|^2 
  \stackrel{(\ref{eq:efield_xfrom})}{=} \frac{2}{c_0\mu_0}|\bm{x}|^2
  \Biggl|\sum^{N_p}_{k=1} \int_{-\infty}^{\infty}\frac{1}{\sqrt{2\pi}}
  \bm{E}_{k}(\bm{x},t_{a})\exp{(j\omega t_{a})}\,dt_{a}\Biggr|^2.
\end{equation}
Substitution of Eq.~\eqref{eq:far_efield} into Eq.~\eqref{eq:far_bunch_inty}, using the approximately constant observation direction $\bm{n}=\bm{x}/|\bm{x}|$, applying the variable transformation given in Eq.~\eqref{eq:far_adv} (with $dt_a = (1-\bm{n}\cdot\bm{\beta}_k'(t))\,dt$ by normalization of $\bm{\beta}_k'=\dot{\bm{x}}_k'/c_{0}$), and setting $\varepsilon_{0}=(\mu_0c_0^2)^{-1}$ in Eq.~\eqref{eq:far_bunch_inty} results in \cite[Eqn.~(14.65)]{jackson_classical_nodate}
\begin{align}
  \frac{dI^2}{d\omega d\Omega}(\bm{n},\omega)
  &\approx\frac{q^2}{16\pi^3\varepsilon_{0}c_{0}}
  \Biggl|\sum^{N_p}_{k=1}
    \int^{\infty}_{-\infty}\underbrace{\frac{\bm{n}\times\left((\bm{n}-\bm{\beta}'_k(t))
    \times{\dot{\bm{\beta}}'_k(t)}\right)}
    {(1-\bm{n}\cdot{\bm{\beta}}'_k(t))^2}}_{\equiv\bm{\mathcal{A}}_{k}(\bm{n}, t)} \frac{1}{(1-\bm{n}\cdot{\bm{\beta}}'_k(t))}\exp(j\omega t_a)\,dt_{a}
  \Biggr|^2\nonumber\\
  &= \frac{q^2}{16\pi^3\varepsilon_{0}c_{0}}
    \Biggl|
      \sum^{N_p}_{k=1} 
      \int^{\infty}_{-\infty}
      \bm{\mathcal{A}}_{k}(\bm{n}, t)
      \exp\left(j\omega \left(t-\dfrac{\bm{n}\cdot \bm{x}'_k(t)}{c_0}\right)\right)\,dt
    \Biggr|^2 \label{eq:const_phase_drop}\\
    & \qquad \qquad \text{with } \quad \bm{\mathcal{A}}_{k}(\bm{n}, t)\equiv
  \frac{\bm{n}\times\left((\bm{n}-\bm{\beta}'_k(t))\times
  \bm{\dot{\beta}}'_k(t)\right)}{(1-\bm{n}\cdot\bm{\beta}'_k(t))^2}.\nonumber
%\\
%&= \dfrac{q^2}{16\pi^3\varepsilon_{0}c_{0}}
%        \left|
%        \sum^{N_p}_{k=1} 
%        \int^{t_{\text{end}}}_{0}
%                \bm{\mathcal{A}}_{k}(\bm{n}, t)
%                \exp\left(j\omega \left(t-\dfrac{\bm{n}\cdot \bm{x}'_k(t)}{c_0}\right)\right)\,dt
%        \right|^2 \label{eq:integral_bound} \\
%&\approx  \dfrac{q^2(\Delta t)^2}{16\pi^3\varepsilon_{0}c_{0}}
%        \left|
%        \sum^{N_p}_{k=1} 
%        \sum^{N_s}_{i=0}
%                \bm{\mathcal{A}}_{k}(\bm{n}, i\Delta t)
%                \exp\left(j\omega \left(i\Delta t-\dfrac{\bm{n}\cdot \bm{x}'_k(i\Delta t)}{c_0}\right)\right)
%        \right|^2, \label{eq:inty_dft}
\end{align}
In Eq.~\eqref{eq:const_phase_drop}, it is the change of variables from $t_a$ to $t$ that circumvents the difficulty to superimpose the electrons' fields at asynchronous future time points.
We have also dropped the common constant phase term $\exp(j\omega|\bm{x}|/c_{0})$ in Eq.~\eqref{eq:const_phase_drop} as it has no impact on the overall amplitude.
%\begin{equation*}
%  \bm{\mathcal{A}}_{k}(\bm{n}, t)\equiv
%  \frac{\bm{n}\times\left((\bm{n}-\bm{\beta}'_k(t))\times
%  \bm{\dot{\beta}}'_k(t)\right)}{(1-\bm{n}\cdot\bm{\beta}'_k(t))^2}.
%\end{equation*}
The time window of the simulation, i.\,e., when the acceleration of the charged particle by the driving field is nonzero, is chosen as $[0,t_{\text{end}}]$. We use an equidistant discretization in time with $N_s$ intervals, $N_s\Delta t=t_{\text{end}}$, leading to
\begin{equation}
    \frac{dI^2}{d\omega d\Omega}(\bm{n},\omega)
    \approx \dfrac{q^2(\Delta t)^2}{16\pi^3\varepsilon_{0}c_{0}}
        \Biggl|
        \sum^{N_p}_{k=1}
        \sum^{N_s}_{i=0}
            \bm{\mathcal{A}}_{k}(\bm{n}, i\Delta t)
            \exp\left(j\omega \left(i\Delta t-\frac{\bm{n}\cdot
            \bm{x}'_k(i\Delta t)}{c_0}\right)\right)
    \Biggr|^2. \label{eq:inty_dft}
\end{equation}

Introducing the abbreviations
\begin{align}
  \label{eq:freq:far_field_freq}
  \bm{\mathcal{E}_{k,i}}(\bm{n},\omega)&\equiv
                \bm{\mathcal{A}}_{k}(\bm{n}, i\Delta t)
                \exp\left(j\omega \left(i\Delta t-\dfrac{\bm{n}\cdot \bm{x}'_k(i\Delta t)}{c_0}\right)\right), \\
	\bm{\mathcal{E}}^\text{bunch}_{i}(\bm{n},\omega) &\equiv \sum^{N_p}_{k=1}\bm{\mathcal{E}}_{k,i}(\bm{n},\omega), \qquad
	\bm{\mathcal{E}}^\text{sum}(\bm{n},\omega)       \equiv \sum^{N_s}_{i=1}\bm{\mathcal{E}}^\text{bunch}_{i}(\bm{n},\omega), \nonumber
\end{align}
and exchanging the order of summation w.\,r.\,t. $k$ and $i$ in \eqref{eq:inty_dft}, we obtain
\begin{align*}
  \dfrac{dI^2}{d\omega d\Omega}(\bm{n},\omega)
  &\approx \frac{q^2(\Delta t)^2}{16\pi^3\varepsilon_{0}c_{0}}
    \Biggl|
        \sum^{N_s}_{i=1} 
        \sum^{N_p}_{k=1}
        \bm{\mathcal{E}}_{k,i}
    \Biggr|^2  \label{eq:inty_dft}
  = \frac{q^2(\Delta t)^2}{16\pi^3\varepsilon_{0}c_{0}}
    \Bigl|
        \sum^{N_s}_{i=1} 
        \bm{\mathcal{E}}^\text{bunch}_{i}
    \Bigr|^2 
  = \frac{q^2(\Delta t)^2}{16\pi^3\varepsilon_{0}c_{0}}
    \left| 
      \bm{\mathcal{E}}^\text{sum}
    \right|^2.
\end{align*}
A pseudocode for the frequency-domain method is provided in Algorithm~\ref{alg:frequency_domain_method}.

\begin{algorithm}[H]
\Init{}{
    $\bm{\mathcal{E}}^\text{sum}(\bm{n},\omega)\leftarrow 0$ \\
}
\For{simulation step $i$}{
    $\bm{\mathcal{E}}^\text{bunch}_{i}(\bm{n},\omega)\leftarrow 0$ \\
    \For{particle $k$}{
        update trajectory\\
        $\bm{\mathcal{E}}^\text{bunch}_{i}(\bm{n},\omega)\leftarrow \bm{\mathcal{E}}^\text{bunch}_{i}(\bm{n},\omega) + \bm{\mathcal{E}}_{k,i}(\bm{n},\omega)$
    }
    $\bm{\mathcal{E}}^\text{sum}(\bm{n},\omega) \leftarrow \bm{\mathcal{E}}^\text{sum}(\bm{n},\omega) + \bm{\mathcal{E}}^\text{bunch}_{i}(\bm{n},\omega)$
}
\caption{Frequency-Domain Method}
\label{alg:frequency_domain_method}
\end{algorithm}

\section{Time-Domain Method}
To start our discussion, we first present the result of a test simulation for $10$ particles initially at rest with random initial positions which are samples from a Gaussian distribution with mean $0$ and standard deviation $1$ in each direction. The particles are driven by a sinusoidal electromagnetic plane wave. In the simulation, the equations of motion for charged particles are simulated using the Boris method~\cite{boris1970}, and the trajectory of each particle is used to evaluate the radiation field arriving at an observation point and the corresponding observation time by evaluating Eq.~\eqref{eq:far_efield} and Eq.~\eqref{eq:far_adv}.

\begin{figure}[h!]
\centering
\begin{subfigure}[b]{0.45\textwidth}
    \includegraphics[width=\linewidth]{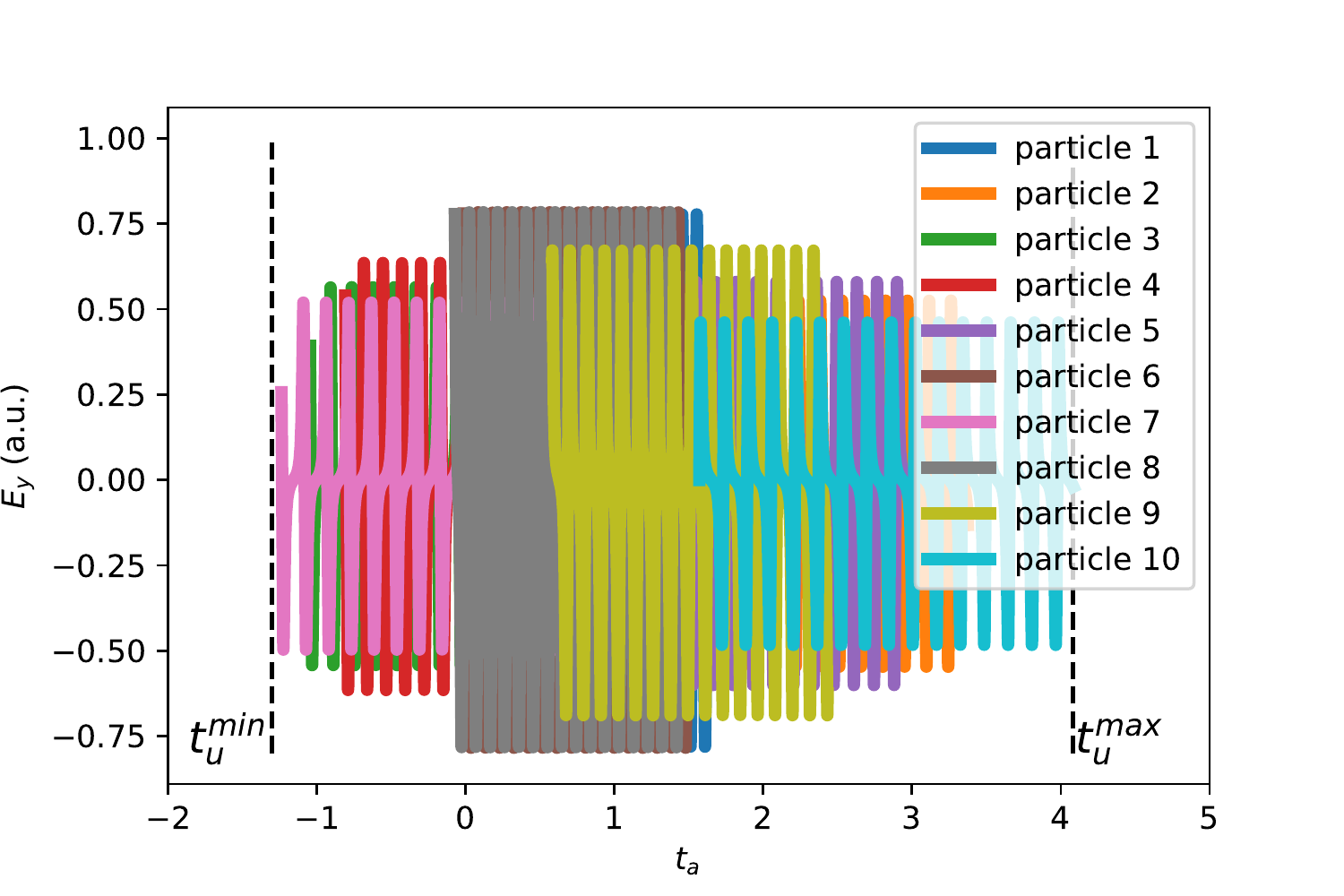}
    \caption{radiation fields of all particles}
\end{subfigure}
\begin{subfigure}[b]{0.45\textwidth}
    \includegraphics[width=\linewidth]{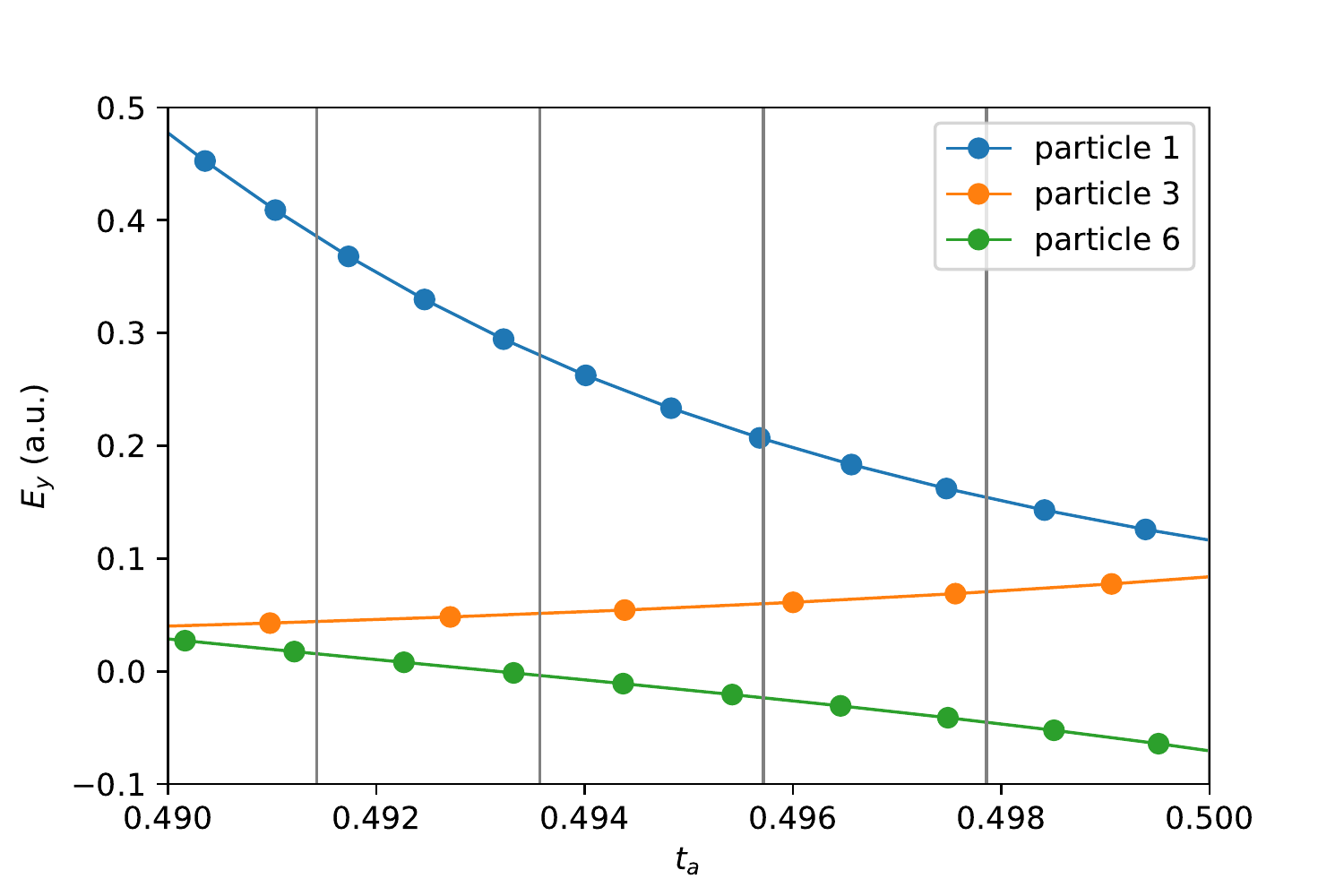}
    \caption{zoom-in for three selected particles}
\end{subfigure}
\caption{Simulation result for the test simulation. (a) shows the temporal profiles of the radiation fields from all particles. (b) shows the zoom-in view of (a) for three selected particles. Here, the relative arrival time of particle radiation field is demonstrated instead of the absolute arrival time.}
\label{fig:advanced_trial}
\end{figure}

The result of this trial simulation is shown in Fig.~\ref{fig:advanced_trial}. Fig.~\ref{fig:advanced_trial}(a) shows the $E_y$ component of the radiation field generated by each particle. Fig.~\ref{fig:advanced_trial}(b) shows a zoom-in view of Fig.~\ref{fig:advanced_trial}(a). From the result of this trial simulation, we  summarize the following observations:
\begin{compactitem}
	\item The observed pulse duration of the radiation fields generated by the particles is different. This is illustrated in Fig.~\ref{fig:advanced_trial}(a) and explained by Eq.~\eqref{eq:far_adv},
	\begin{equation*}
		\int dt^{k}_{a}=\int^{t_{\text{end}}}_{t_{\text{start}}} \left(1-\bm{n}\cdot\bm{\beta}'_{k}(t)\right)\,dt=t^{k}_{a}(t_{\text{end}})-t^{k}_{a}(t_{\text{start}}).
	\end{equation*}
	Each particle moves at different phase of the external electromagnetic wave, and the  particle velocities $\bm{\beta}'_k(t)$ driven by the external field during the simulation can be different. Therefore, the resulting integrals of $dt^k_a$ are different for the particles.
	\item It is problematic to superimpose the fields of all particles since the radiation fields from the trajectories of different particles evaluated at the same time $t$ are observed at different future times $t_a^k=\approx t +|\bm{x}|/c_0-(\bm{n}\cdot \bm{x}'_{k}(t)) / c_0$ \eqref{eq:far_adv}. This is illustrated in Fig.~\ref{fig:advanced_trial}(b).
	\item To superimpose the radiation fields, we define a uniform time grid $\bm{t}_{u}$ (indicated by the gray vertical lines in Fig.~\ref{fig:advanced_trial}(b) and interpolate the particles' fields to the uniform grid (details on the interpolation will follow in subsection \ref{sec:interpolation}).
	\item To cover the radiation temporal profile of all particles, we need to determine the upper and lower bound (annotated by $t^{\text{max}}_{u}$ and $t^{\text{min}}_u$ in Fig.~\ref{fig:advanced_trial}(a) of the uniform time grid $\bm{t}_{u}$. 
\end{compactitem}

%\subsection{Determine the lower/upper bound of the uniform time grid}
To create a uniform time grid for the radiation field interpolation, we need to know the bound of the radiation temporal profile at different observation positions. For a simulation with $N_p$ particles and $N_{\text{obs}}$ observers, we determine the bounds of the uniform (advanced) time grid by
\begin{align}
	t^{\text{min}}_u &= \min\left\{t^{k,m}_{a}(t_{\text{start}}) \mid k=1,\dots,N_p, ~m=1,\dots,N_{\text{obs}}\right\},\label{eq:tu_min}\\
	t^{\text{max}}_u &= \max\left\{t^{k,m}_{a}(t_{\text{end}}) \mid k=1,\dots,N_p, ~m=1,\dots,N_{\text{obs}}\right\},\label{eq:tu_max}
\end{align} 
where $t_{\text{start}}, t_{\text{end}}$ denote the start/end of the simulation and
$$t^{k,m}_a (t) \approx t +\dfrac{|\bm{x}_m|}{c_0}-\dfrac{\bm{n}_m\cdot \bm{x}'_{k}(t)}{c_0}.$$
($|\bm{x}_m|$ is the same for all $m$ observers since they are located on a spherical surface.)
The lower bound $t^{\text{min}}_{u}$ can be easily determined at the beginning of the simulation since we already have the initial positions of all particles. To determine the upper bound $t^{\text{max}}_{u}$, it might be possible to perform a trial simulation in which we only compute the particles' trajectories and evaluate $t^{\text{max}}_{u}$ by Eq.~\eqref{eq:tu_max} at the end of trial simulation. However, this will become costly when $N_p$ is large. To avoid such a trial simulation, %which introduces additional computation on solving particle trajectories, 
we determine $t^{\text{max}}_{u}$ by estimating the total radiation pulse duration from a particle bunch as follows.

\begin{figure}
    \centering
    \includegraphics{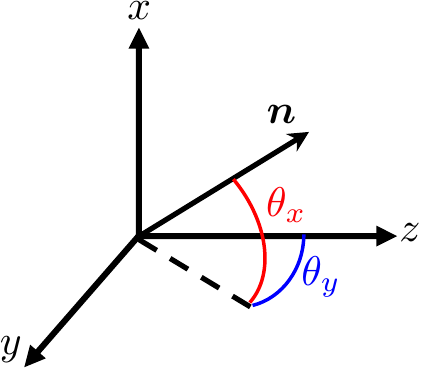}
    \caption{Definition of the normal vector $\bm{n}$ pointing from the particle to the observer. $\theta_x$ is the angle between $\bm{n}$ and its projection to the y-z plane. $\theta_y$ is the angle between the z-axis and the projection of $\bm{n}$ to the z-y plane.}
    \label{fig:nvector}
\end{figure}

As can be observed from Fig.~\ref{fig:advanced_trial}(a), the total radiation pulse duration from all particles is influenced by the radiation pulse durations of single particles and the different lags in their arrival times. For the ICS problem, the radiation pulse duration from a single particle
\begin{equation*}
    T_{\text{rad}}\equiv \int dt_{a}=\int^{T_{\text{laser}}}_{0}\left(1-\bm{n}\cdot\bm{\beta}'(t)\right)\,dt,
\end{equation*}
can be approximated for highly relativistic particles by
\begin{equation}
    T_{\text{rad}}\approx \dfrac{T_{\text{laser}}}{2\gamma^2}
    \label{eq:Tu_second}
\end{equation}
where $T_{\text{laser}}=t_{\text{end}}-t_{\text{start}}$ is the pulse duration of the counter propagating laser pulse. The maximum difference in arrival times of the radiation pulses from different particles is caused by their distribution in space and can be estimated by the last term of Eq.~\eqref{eq:far_adv}, 
\begin{equation}
    T^{\text{bunch}}_{\text{rad}} \equiv
    %\max_{\substack{i,j = 1,\ldots,N_p\\\bm{n}}}
    \max_{\theta_x,\theta_y}\max_{i,j = 1,\ldots,N_p}
    \left|\dfrac{\bm{n}\cdot(\bm{x}'_i - \bm{x}'_j)}{c_0}\right|,
    \label{eq:max_difference_arrival_time}
%\end{equation}
\quad\text{where}\quad 
%\begin{equation*}
    \bm{n}\equiv (\sin\theta_x,\cos\theta_x\sin\theta_y,\cos\theta_x\cos\theta_y).
\end{equation}
$\theta_x$ and $\theta_y$
%\done{JZ: are these the vectors of angles of all observers? Or are these scalar quantities like in the preceeding equation? Then they should not be set in bold.} 
are the angles defining the observers' positions on the sphere (see Fig.~\ref{fig:nvector}). In the highly relativistic scenario, the opening angle of the radiation from a particle is of the order of $1/\gamma\ll1$. We hence use approximations $\sin \theta\approx \theta$, $\cos\theta\approx 1$, and Eq.~\eqref{eq:max_difference_arrival_time} becomes
\begin{equation*}
    T^{\text{bunch}}_{\text{rad}} \approx
    \max_{\substack{i,j = 1,\ldots,N_p\\ |\theta_x|,\,|\theta_y|<1/\gamma}} 
    \left|
        \dfrac{\theta_{x} (x'_{i}-x'_{j}) + \theta_{y}(y'_{i}-y'_{j}) + (z'_{i}-z'_{j})}{c_{0}}
    \right|,
\end{equation*}
which has the upper bound
\begin{equation}
    T^{\text{bunch}}_{\text{rad}} \lessapprox
    \dfrac{l_{b,x}}{\gamma c_0} + \dfrac{l_{b,y}}{\gamma c_0} + \dfrac{l_{b,z}}{c_0},
    \label{eq:Tu_first}
\end{equation}
where $l_{b,x}$, $l_{b,y}$ and $l_{b,z}$ denote the size of the particle bunch in each direction at $t_{\text{start}}$.

Thus, to cover the total radiation pulse duration from all particles, the length of the required uniform time grid $T_u$ can  be approximately chosen as  
\begin{equation}
    T_{u} \approx T^{\text{bunch}}_{\text{rad}} + T_{\text{rad}} \approx \dfrac{l_{b,x}}{\gamma c_0} + \dfrac{l_{b,y}}{\gamma c_0} + \dfrac{l_{b,z}}{c_0} + \dfrac{T_{\text{laser}}}{2\gamma^2}.
\label{eq:time:Tu_estimation}
\end{equation}
Table~\ref{tb:Tu_estimation_cases} lists the values for $T_u$ computed from Eq.~\eqref{eq:time:Tu_estimation} for different parameter settings. 
We also provide values for $T_u$ from a trial simulation to verify the theoretical estimates.
Once $t^{\text{min}}_{u}$ and $T_{u}$ are known, $t^{\text{max}}_{u}=t^{\text{min}}_{u}+T_u$ can be immediately determined.

\begin{table}\renewcommand{\arraystretch}{1}
\centering
\begin{tabular}{cccccll@{\hspace*{2em}}l@{\hspace*{3em}}l}\hline\hline
$l_{b,x}$ & $l_{b,y}$& $l_{b,z}$ & $T_{\text{laser}}$& $\gamma$ & $T^{\text{bunch}}_{\text{rad}}$ ~\eqref{eq:Tu_first} & $T_{\text{rad}}$ ~\eqref{eq:Tu_second} & \hspace*{-.7em}$T_u$ (theory) & \hspace*{-1.6em}$T_u$ (simulation)\\
\hline
 10  & 10 & 10 & 10 & 40 & 35.0000 & 0.003125 & 35.0031 & 35.0211 \\ 
 20  & 10 & 10 & 10 & 40 & 35.8333 & 0.003125 & 35.8365 & 35.8876 \\
 10  & 20 & 10 & 10 & 40 & 35.8333 & 0.003125 & 35.8365 & 35.8540 \\
 10  & 10 & 20 & 10 & 40 & 68.3333 & 0.003125 & 68.3365 & 68.3328 \\ 
 10  & 10 & 10 & 20 & 40 & 35.0000 & 0.006250 & 35.0063 & 35.0310 \\
\hline
\end{tabular}\renewcommand{\arraystretch}{1}
\caption{Theoretical estimation of $T_u$ by Eq.~\eqref{eq:time:Tu_estimation} for different parameter settings in comparison with results from trial simulations for the particles' trajectories using Eqs.~\eqref{eq:tu_min}, Eq.~\eqref{eq:tu_max}. Here, all length and time quantities are normalized to $2\pi/\lambda_0$ and $2\pi c_0/\lambda_0$, respectively, with $\lambda_0 = 1000 \mu m$.
%The values computed for each parameter setting is also provided for comparison with the analytic prediction. From the simulation using the particle's trajectory from the beginning and the end, $t^{\text{max}}_u$ and $t^{\text{min}}_u$ (hence $T_u$) can be directly computed by Eq.~\eqref{eq:tu_max} and Eq.~\eqref{eq:tu_min}.
}
\label{tb:Tu_estimation_cases}
\end{table}

\subsection{Interpolation}\label{sec:interpolation}
In the previous subsection, we discussed how to choose the bounds of a uniform time grid in order to determine the temporal positions for the computation of the total radiation field. Next, we interpolate the particles' fields at different temporal (advanced) positions to the uniform grid by piecewise linear interpolation.
% justification of interpolation scheme is commented out
\begin{comment}
We use piecewise linear interpolation for the following reasons:
\begin{compactenum}
	\item It is easy and robust.
	\item Higher order methods would require additional points. This can be memory-demanding since we have to store the field at several temporal positions for each particle. Therefore, the linear interpolation method is advantageous in view of memory requirements because only two points are needed for the interpolation.
	\item In the spectrum calculation, the maximum radiation frequency resolved by the simulation is determined by the average of the spacing of radiation arrival time $\pi/\langle\Delta t_a\rangle$ \cite{pausch_electromagnetic_2012}. 
	To observe the maximum radiation frequency achieved by the simulation, the spacing of the uniform grid $\Delta t_u$ should be chosen the same as $\langle\Delta t_a\rangle$. 
	Under this condition, there will be only one grid point between two successive points of radiation arrival time. The piecewise linear interpolation may be enough for accuracy.\todo{I don't get this argument...why do we believe linear interpolation is enough?}
\end{compactenum}
\end{comment}
\begin{figure}[ht]
    \begin{subfigure}[b]{0.45\textwidth}
        \includegraphics[width=\textwidth]{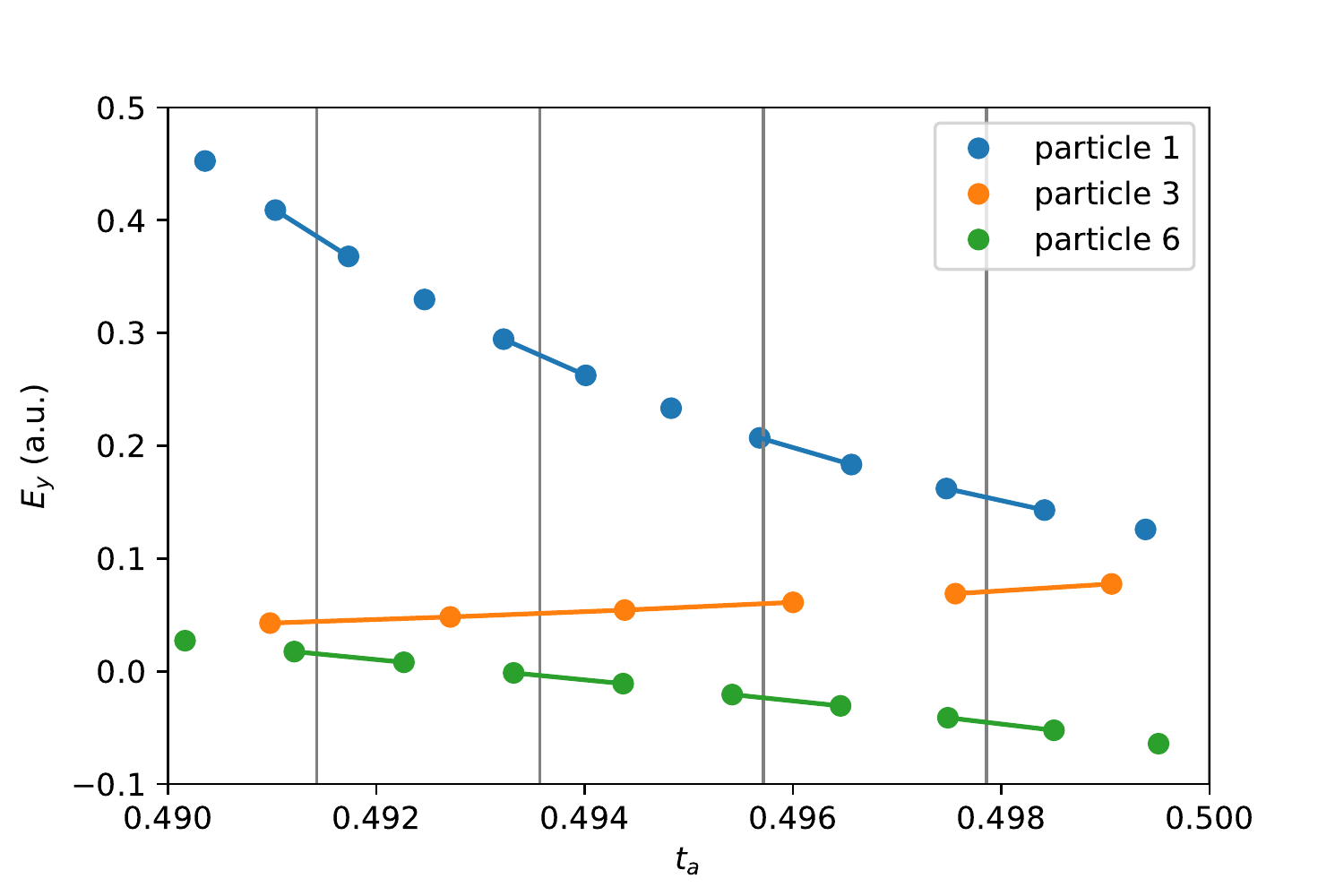}
        \caption{local linear interpolation}
    \end{subfigure}
	\begin{subfigure}[b]{0.45\textwidth}
        \includegraphics[width=\textwidth]{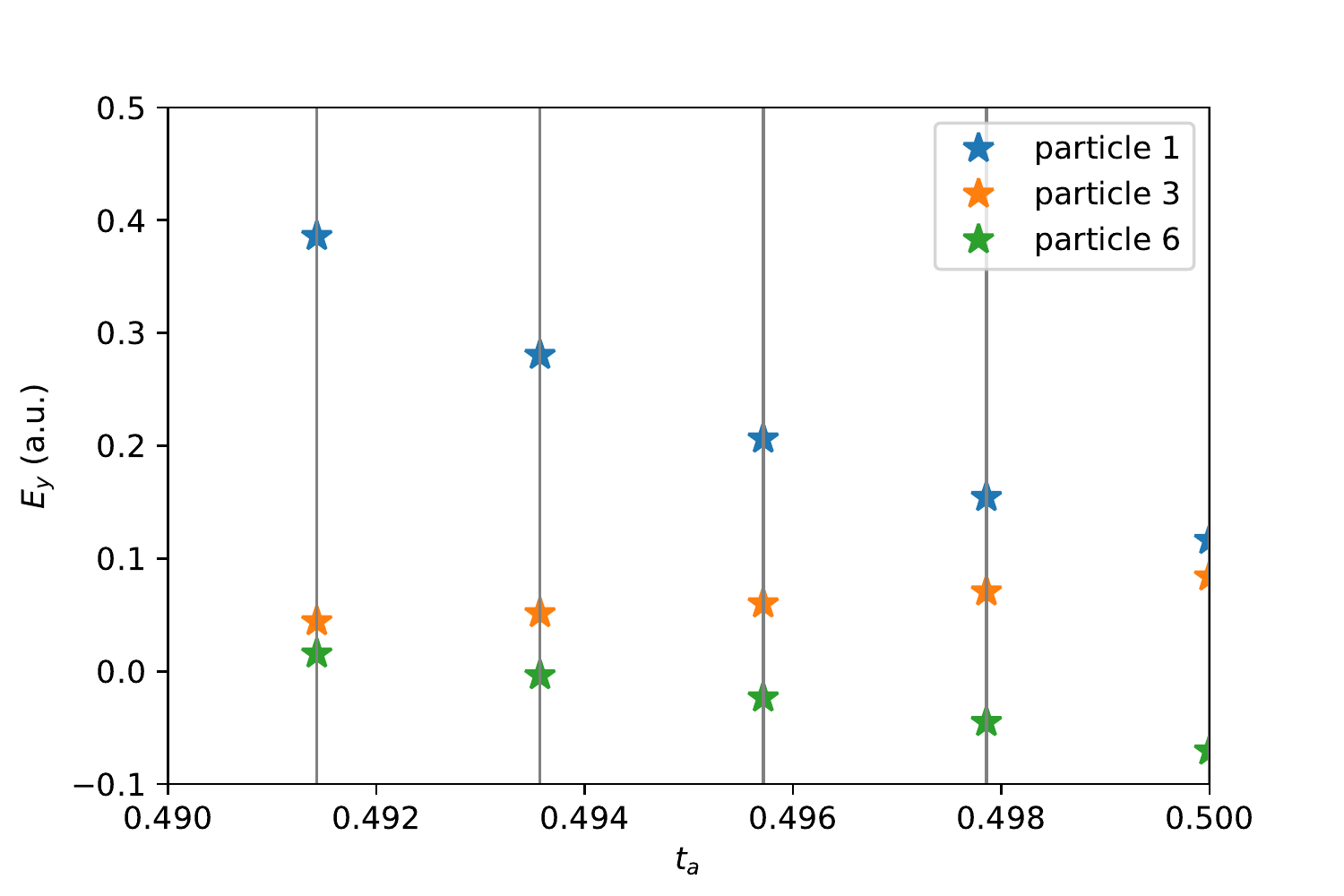}
        \caption{resulting grid values}
    \end{subfigure}
	\caption{Linear interpolation scheme for the temporal radiation profile. The gray vertical lines mark the position of the uniform time grid points. The round points in (a) connected by line segments are used for interpolation. (b) shows the result after interpolation. The stars indicate the values of the interpolated field.}
	\label{fig:time:interpolation}
\end{figure}

To apply linear interpolation, one needs the arrival times and corresponding radiation fields from each particle at the previous and current time steps. 
This might suggest that we need two memory buffers for storing the field information of adjacent time steps. 
However, we actually only need one memory buffer to store the field information of the previous step. 
In every time step, we compute the field in the current time step for each particle and interpolate with the field stored in the memory buffer. Once the interpolated field and the corresponding position at the uniform time grid are calculated, the interpolated field is superimposed to the value at the array for storing the total radiation field of the particles. The complete procedure is described in Algorithm~\ref{alg:time_domain_method} (for a single observation direction $\bm{n}$).

\begin{algorithm}[H]
\Init{}{
    $\bm{E}_{tot}(\bm{n}, t_u) \leftarrow 0$ (total field from all particles over a uniform grid with points $t_u$)\\
    $t^k_b(\bm{n})\leftarrow 0$ (buffer for storing of the field's arrival time of particle $k$ computed at previous step) \\
    $\bm{E}^{k}_{b}(\bm{n})\leftarrow 0$ (buffer for storing the arrival field of particle $k$ computed at previous time step)
}
\For{simulation step $i$}{
    \For{particle $k$}{
        update trajectory\\
        compute $t^k_a(\bm{n})$ and $E^{k}(\bm{n}, t^k_a)$\\
        \For{$t_u:\,t^k_b(\bm{n}) < t_u \le t^k_a(\bm{n})$}{
        $\bm{E}_{tot}(\bm{n}, t_u) \leftarrow \bm{E}_{tot}(\bm{n}, t_u) +
            E^{k}_{b}(\bm{n})\tfrac{t^k_{a}(\bm{n}) - t_u}{t^k_a(\bm{n})-t^k_b(\bm{n})}
            +
            \bm{E}^{k}(\bm{n}, t^k_a)
            \tfrac{t_u-t^k_b(\bm{n})}{t^k_a(\bm{n})-t^k_b(\bm{n})}$
        }
    }
    $t^k_b(\bm{n}) \leftarrow t^k_a(\bm{n})$ \\
    $\bm{E}^{k}_{b}(\bm{n}) \leftarrow \bm{E}^{k}(\bm{n}, t^k_a)$
}
\caption{Proposed Time-Domain Method}
\label{alg:time_domain_method}
\end{algorithm}

\subsection{Parallelization}
Before discussing parallelization, we briefly introduce key data structures used in the code development. Here, the memory cost is measured by the number of floating-point numbers to be stored.
\begin{description}
	\item[Beam:] A structure to store the state of motion (position, momentum, velocity and acceleration) for a bunch of particles at a specific time. For a particle bunch with $N_p$ particles in three dimensional space, $12N_p$ floating-point numbers are required.
	\item[Sensor:] A sensor records the total radiation field from a bunch of particles over the uniform time grid. For $N_{T_u}$ uniform time grid points, $3N_{T_u}$ floating-point numbers are required. In addition, each particle's radiation arrival time and radiation field are also stored in a memory buffer which requires $4N_p$ floating-point numbers. 
	\item[Detector:] A detector contains $N_{\text{obs}}$ sensors and each sensor has a different observation position. The memory cost for the detector is thus  $N_{\text{obs}}(4N_p + 3N_{T_u})$ floating-point numbers.
\end{description}

\begin{figure}[th]
\centering % width as .8*w_i/(w_1+w_2) in pixels, i=1,2
\begin{subfigure}[t]{.38\textwidth}
    \vskip 0pt
    \includegraphics[width=\linewidth]{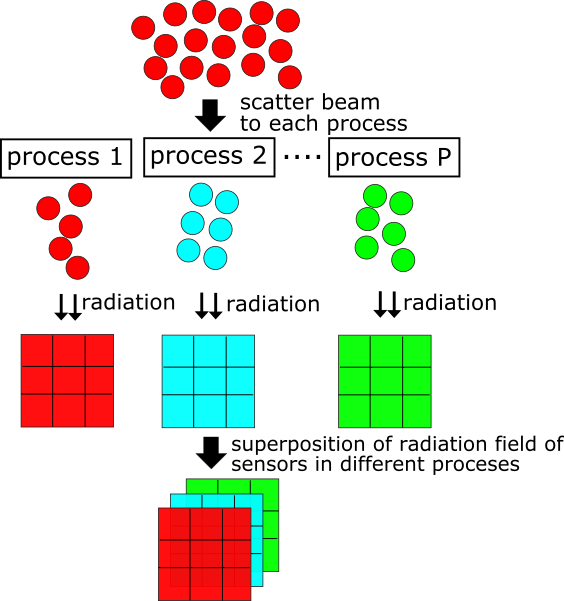}
    \caption{beam parallelization}
\end{subfigure}\hspace*{.5cm}
\begin{subfigure}[t]{.419882\textwidth}
    \vskip 0pt
    \includegraphics[width=\linewidth]{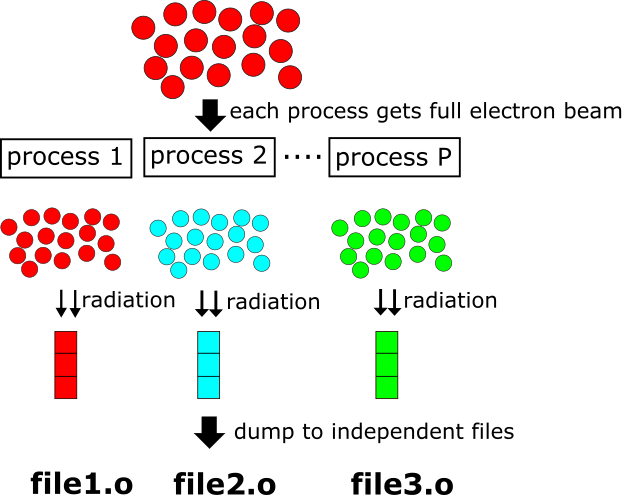}
    \caption{detector parallelization}
\end{subfigure}
\caption{Two possible parallelization schemes: (a) beam parallelization and (b) detector parallelization.}
\label{fig:parallelization_scheme}
\end{figure}

There are at least two possible parallelization schemes for the simulation, see Fig.~\ref{fig:parallelization_scheme}. 
One is beam parallelization in which the particle bunch is divided into several small bunches and each small bunch is assigned to an MPI process.
Each process creates its own detector object which receives the radiation from a small bunch. 
At the end of the simulation, the radiation data in each detector object is superimposed to the radiation data in the detector object created by the master task. 
The other possible scheme is detector parallelization where a global sensor in the detector is divided into several local detectors. 
In the beginning of the simulation, each process creates a copy of the entire electron bunch and a local detector consisting of a subset of sensors from the global detector. 
During the simulation, the local detector receives the radiation from the copy of the entire electron bunch owned by each process. 
At the end of the simulation, each process dumps the data from its local detector separately.

The simulation of ICS can involve a large number of particles and sensors.
Therefore, the memory requirement per process for both parallelization schemes is an important consideration. 
For beam parallelization, the memory cost per process is
\begin{equation*}
	M_{\text{beam}}(P, N_p,N_{\text{obs}},N_{T_u}) = 12\tfrac{N_p}{P}+N_{\text{obs}}\left(4\tfrac{N_p}{P}+3N_{T_u}\right)
\end{equation*}
where $P$ is the number of computer processes. 
For detector parallelization, the memory cost per process is
\begin{equation*}
	M_{\text{detector}}(P, N_p,N_{\text{obs}},N_{T_u}) = 12N_p+\tfrac{N_{\text{obs}}}{P}\left(4N_p+3N_{T_u}\right).
\end{equation*}
In order to determine which scheme has lower memory cost, we define the function 
\begin{align}
\phi(P, N_p,N_{\text{obs}},N_{T_u}) &\equiv
\dfrac{M_{\text{beam}}(P, N_p,N_{\text{obs}},N_{T_u})}{M_{\text{detector}}(P, N_p,N_{\text{obs}},N_{T_u})} = 
\dfrac{12\tfrac{N_p}{P}+N_{\text{obs}}\left(4\tfrac{N_p}{P}+3N_{T_u}\right)}{12N_p+\tfrac{N_{\text{obs}}}{P}\left(4N_p+3N_{T_u}\right)}\nonumber\\
 &=1+\dfrac{3\left(1-\tfrac{1}{P}\right)\left(N_{\text{obs}}N_{T_u}-4N_p\right)}{12N_p+\tfrac{N_{\text{obs}}}{P}\left(4N_p+3N_{T_u}\right)}.
\label{eq:time:parallel_scheme_function}
\end{align}
If $\phi>1$, detector parallelization has a lower memory footprint and otherwise beam parallelization.
Since the factor $(1-1/P)$ in the enumerator and all terms in the denominator of the second term are positive, we conclude that
\begin{equation}
\phi(P,N_p,N_{\text{obs}},N_{T_u})\quad 
\begin{cases}
\ge 1 \quad \text{ if } N_{\text{obs}}N_{T_u}-4N_p \ge 0, \\
< 1 \quad \text{ else. }	
\end{cases}\label{eq:time:parallel_scheme_function_2}
\end{equation}
Therefore, the ratio $N_{\text{obs}}N_{T_u}/(4N_p)$ determines which parallelization scheme is more favorable in terms of memory consumption.

The number of particles $N_p$ can be computed from the total charge of the electron bunch. 
The number of uniform grid points $N_{T_u}$ should be chosen according to the Nyquist theorem
\begin{equation}
    N_{T_u}=2\dfrac{\omega_{\text{max}}}{\Delta \omega}. \\
    \label{eq:NTu_estimation_1}
\end{equation}
Here, $\omega_{\text{max}}$ is the maximum radiation angular-frequency we want to observe which can be estimated by
\begin{equation}
    \omega_{\text{max}}=4\gamma^2\omega_0
    \label{eq:NTu_estimation_2}
\end{equation}
for a typical ICS problem where $\omega_0$ is the angular-frequency of the counter-propagating laser. The resolution for the angular-frequency is determined by the duration of the total radiation pulse from particles
\begin{equation}
    \Delta \omega = \dfrac{2\pi}{T_u}.
    \label{eq:NTu_estimation_3}
\end{equation}
Combining Eq.~\eqref{eq:NTu_estimation_1}, Eq.~\eqref{eq:NTu_estimation_2}, Eq.~\eqref{eq:NTu_estimation_3} and Eq.~\eqref{eq:time:Tu_estimation}, we can estimate 
\begin{equation*}
    N_{T_u}
    \approx 8\gamma^2\left(\dfrac{l_{b,x}}{\gamma \lambda_0} + 
        \dfrac{l_{b,y}}{\gamma\lambda_0} + \dfrac{l_{b,z}}{\lambda_0} + \dfrac{c_0 T_{\text{laser}}}{2\gamma^2\lambda_0}\right)
\end{equation*}
where the identity $\omega_0=2\pi c_0/\lambda_0$ is used.

In Table~\ref{tb:Ntu_Ns_ratio_for_ics_experiments}, we use parameters from different experimental projects of ICS sources to compute the ratio $4N_{T_p}/N_{T_u}$ which is the number of observers where both parallelization schemes break even ($\phi=1$). Hence, according to \eqref{eq:time:parallel_scheme_function_2}, for a larger number of observers, detector parallelization is preferable whereas a smaller number of observers should be computed with beam parallelization.  
In general, we need at least a few thousand observation angles ({i.\,e.}, $N_{\text{obs}}>1000$) to sufficiently resolve the radiation angular distribution. 
Therefore, detector parallelization is more favorable and is thus  implemented in the solver. 

\subsection{Implementation}
\begin{figure}[h!]
    \centering
    \begin{subfigure}[b]{0.45\textwidth}
        \includegraphics[width=0.9\textwidth]{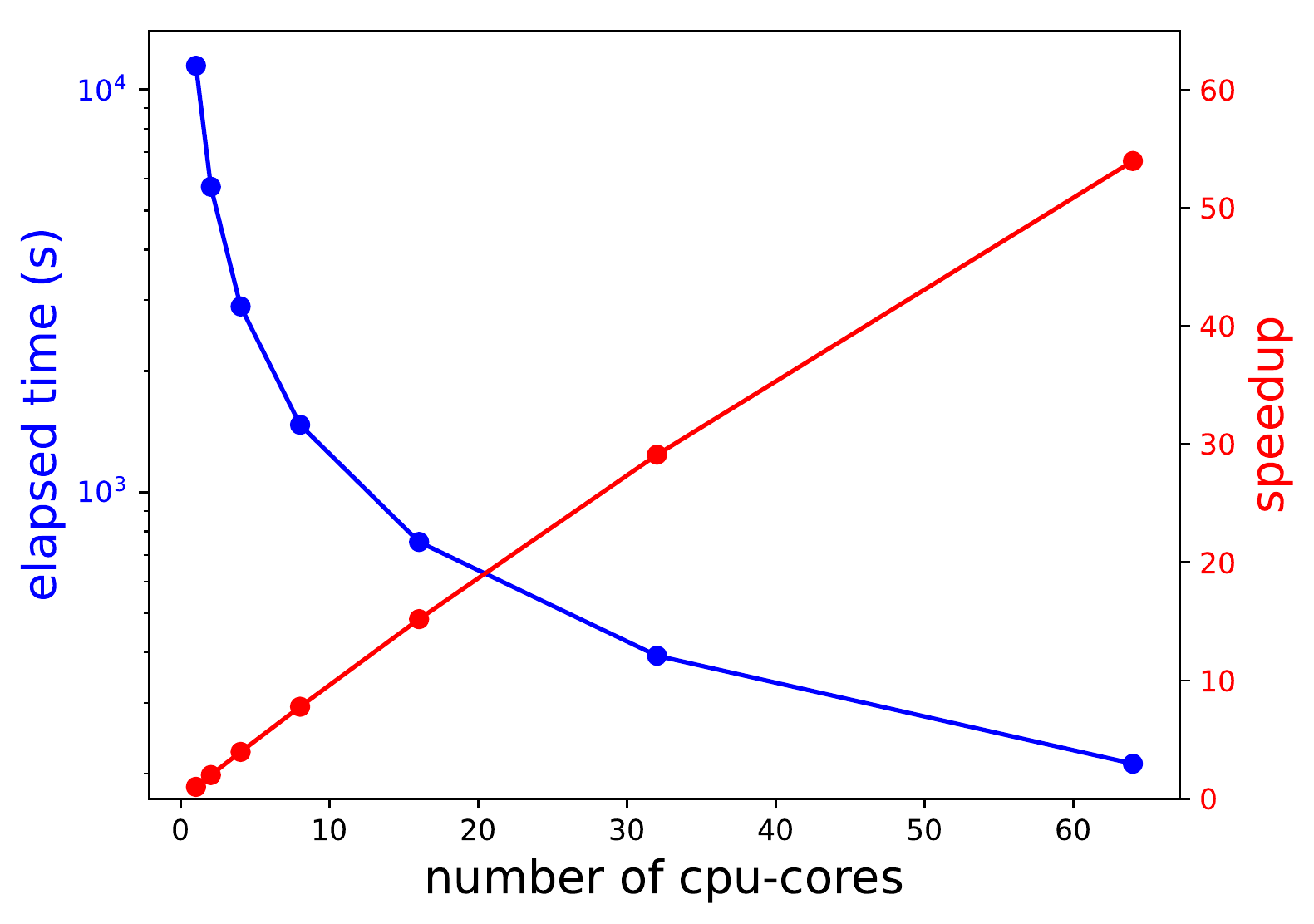}
        \caption{Time-domain method}
    \end{subfigure}
    \begin{subfigure}[b]{0.45\textwidth}
        \includegraphics[width=0.9\textwidth]{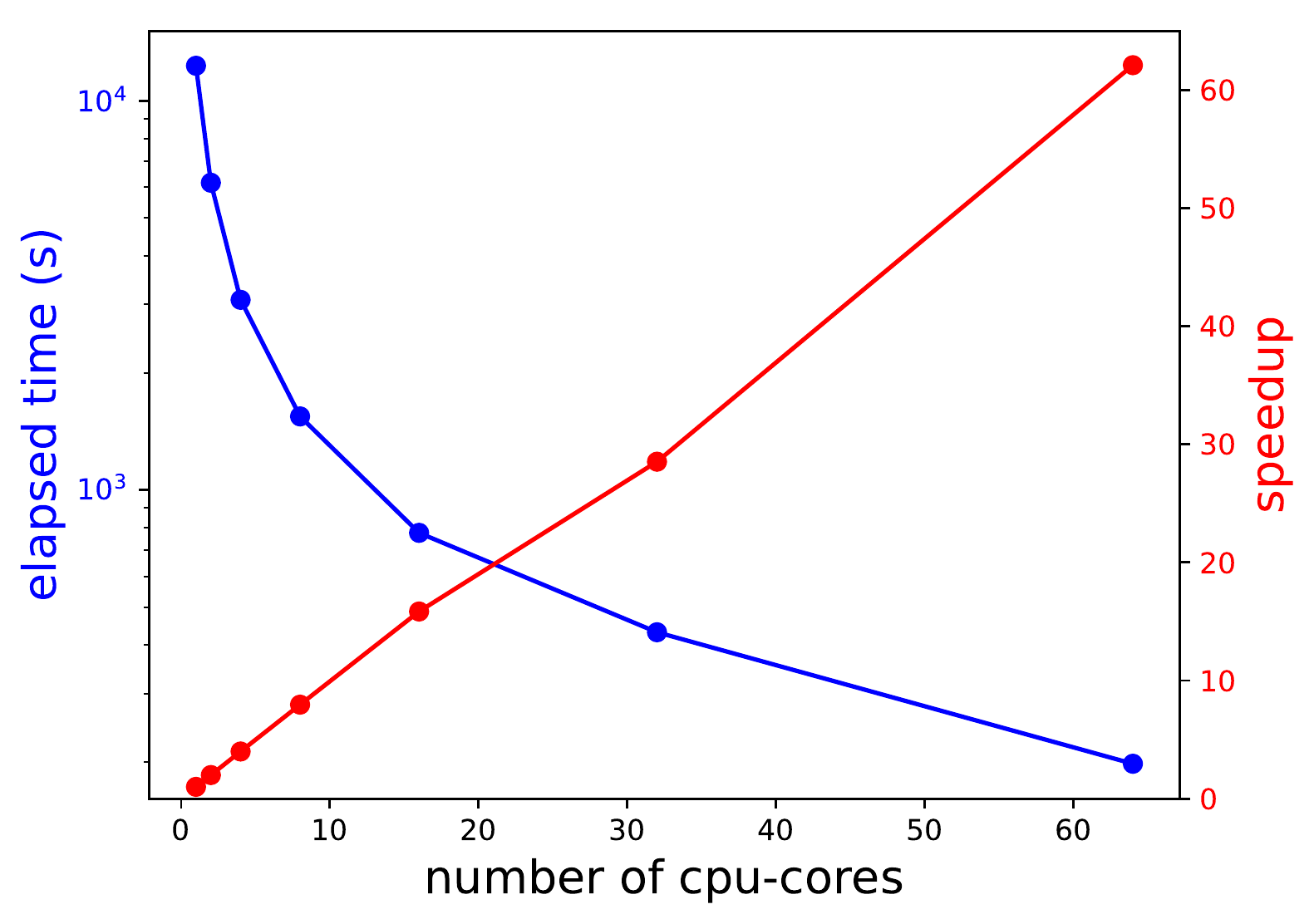}
        \caption{Frequency-domain method}
    \end{subfigure}
    \caption{Parallel performance for: (a) proposed time-domain simulation with $N_p=10^5$, $N_{\text{obs}}=1024$ and $N_{T_u}=10^4$; (b) frequency-domain simulation with $N_p=1024$, $N_{\text{obs}}=64$ and $N_{\omega}=500$.}
    \label{fig:parallelization_performance}
\end{figure}

We use C++ and MPI to implement the time-domain algorithm and the detector parallelization scheme described in this section. The parallel performance for the code is demonstrated in Fig.~\ref{fig:parallelization_performance}(a). For comparison, the parallel performance of our implementation of the frequency-domain method is also given in Fig.~\ref{fig:parallelization_performance}(b). A parallelized post-processing code written in Julia and MPI.jl is used to transform the time-domain field to the radiation spectra. The two solvers can be accessed via
\url{https://doi.org/10.5281/zenodo.5139340}.
%\done{for a paper like this, I would strongly argue in favour of publishing the code with it unless there are copyright issues at play. if we decide to publish it, create a release on github and assign it a doi via Zenodo (see \url{https://guides.github.com/activities/citable-code/}) and cite it in the paper}

%\begin{table}\renewcommand{\arraystretch}{1.1}
%\centering
%\begin{tabular}{lcccccc}\hline\hline
%\multicolumn{1}{c}{project name}   & beam energy (MeV) & $l_z$ (\SI{}{\micro\metre})& charge (pC)&  $T_{\text{laser}}$ (ps) & $\lambda_0$ (\SI{}{\micro\metre}) & $N_{T_u}/N_{p}$ \\
%\hline
%AXSIS \cite{kartner_axsis_2016} & 19.5  & 2.9   & $1$ & 1    & 1    &   $ 5.60 \times 10^{-3}$ \\ 
%ODU CLS \cite{deitrick_inverse_2017, deitrick_high-brilliance_2018}& 25  & 382   & 10    & 1.57           & 1             &    0.117   \\
%ASU CXFEL \cite{graves_asu_2018}& 35    & 1.5        &  1               & 1.5               & 1.03  & $9.02\times 10^{-3}$    \\
%ASU CXLS \cite{graves_compact_2014}& 40    & 147        &  100               & 3.0              & 1.03              &   $1.12\times 10^{-2}$   \\
%XFI \cite{brummer_design_2020}& 65.9     &       3     & 10                 & 1.7             & 0.8               &   $8.02\times 10^{-3}$  \\ 
%MuLCS \cite{gunther_versatile_2020}& 45          & 15000      & 1000               & 26              & 1.064            &    0.140    \\ 
%ThomX \cite{variola_thomx_nodate,dupraz_thomx_2020}& 70          & 6000       & 1000               & 11.75           & 1.03             &    0.140   \\
%\hline
%\end{tabular}
%\caption{The $N_{T_u}/N_{p}$ ratio for different experimental projects. Here, the electron beam transverse sizes $l_x$ and $l_y$ are not considered in the evaluation of $N_{Tu}$ since their contribution is minor for a high energy electron beam. }
%\label{tb:Ntu_Ns_ratio_for_ics_experiments}
%\end{table}

\begin{table}\renewcommand{\arraystretch}{1.1}
\centering
\begin{tabular}{lcccccc}\hline\hline
\multicolumn{1}{c}{project name}   & beam energy (MeV) & $l_z$ (\SI{}{\micro\metre})& charge (pC)&  $T_{\text{laser}}$ (ps) & $\lambda_0$ (\SI{}{\micro\metre}) & $4N_{p}/N_{T_u}$ \\
\hline
AXSIS \cite{kartner_axsis_2016} & 19.5  & 2.9     & $1$ & 1  & 1   &   $714.3$       \\ 
ODU CLS \cite{deitrick_inverse_2017, deitrick_high-brilliance_2018}& 25        & 382       & 10              & 1.57           & 1             &   34.2   \\
ASU CXFEL \cite{graves_asu_2018}& 35    & 1.5        &  1               & 1.5               & 1.03  & $443.5$    \\
ASU CXLS \cite{graves_compact_2014}& 40    & 147        &  100               & 3.0              & 1.03              &   $357.1$   \\
XFI \cite{brummer_design_2020}& 65.9     &       3     & 10                 & 1.7             & 0.8               &   $498.8$  \\ 
MuLCS \cite{gunther_versatile_2020}& 45          & 15000      & 1000               & 26              & 1.064            &    28.6    \\ 
ThomX \cite{variola_thomx_nodate,dupraz_thomx_2020}& 70          & 6000       & 1000               & 11.75           & 1.03             &    28.6   \\
\hline
\end{tabular}
\caption{The ratio $4N_{p}/N_{T_u}$ ratio for different experimental projects. Here, the electron beam transverse sizes $l_x$ and $l_y$ are not considered in the evaluation of $N_{T_u}$ since their contribution is minor for a high energy electron beam. }
\label{tb:Ntu_Ns_ratio_for_ics_experiments}
\end{table}
\section{Numerical comparison of the time-domain and frequency-domain methods}

\subsection{Accuracy}\label{sec:comparison:accuracy}
To compare the accuracy of the radiation spectra computed by both methods, we measure the component-wise relative error 
\begin{equation}
    \text{error}(\omega, \theta) = \text{abs}\left(\dfrac{dI}{d\omega d\Omega}\bigg|_\text{simulation} - \dfrac{dI}{d\omega d\Omega}\bigg|_\text{theory}\right) / \max\left(\dfrac{dI}{d\omega d\Omega}\bigg|_\text{theory}\right).
\label{eq:spectra_error}
\end{equation}
between the theoretical and computed results~\cite{pausch_how_2014}.
The theoretical result is computed by the formula proposed by Esarey et al.~\cite{esarey_nonlinear_1993} which considers the radiation spectral-angular distribution from a single particle interacting with a finite number of periods of a sinusoidal electromagnetic wave. 

\begin{figure}
\begin{subfigure}[b]{0.33\linewidth}
        \includegraphics[width=\textwidth]{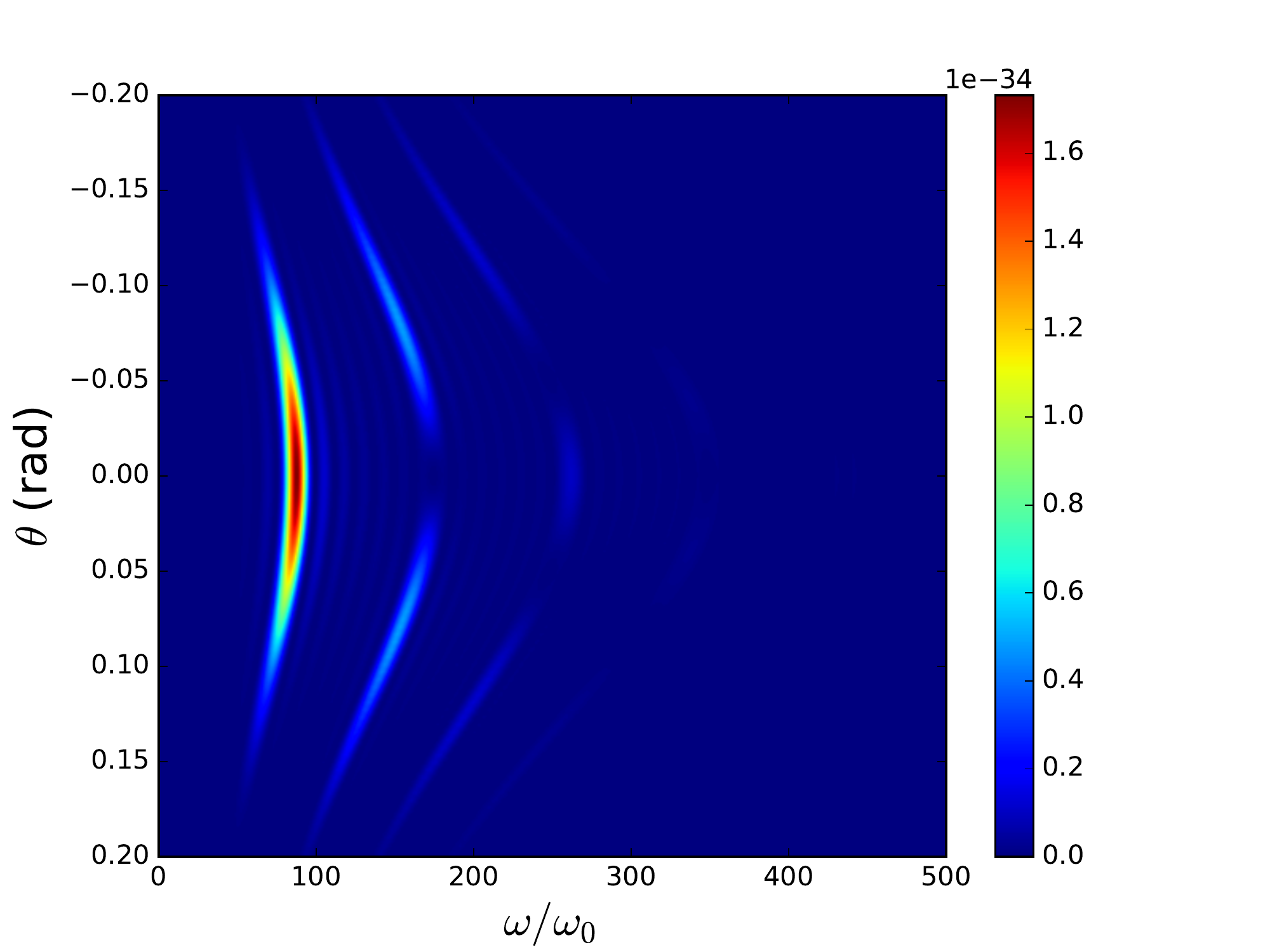}
    \caption{theory}
\end{subfigure}
\begin{subfigure}[b]{0.33\linewidth}
        \includegraphics[width=\textwidth]{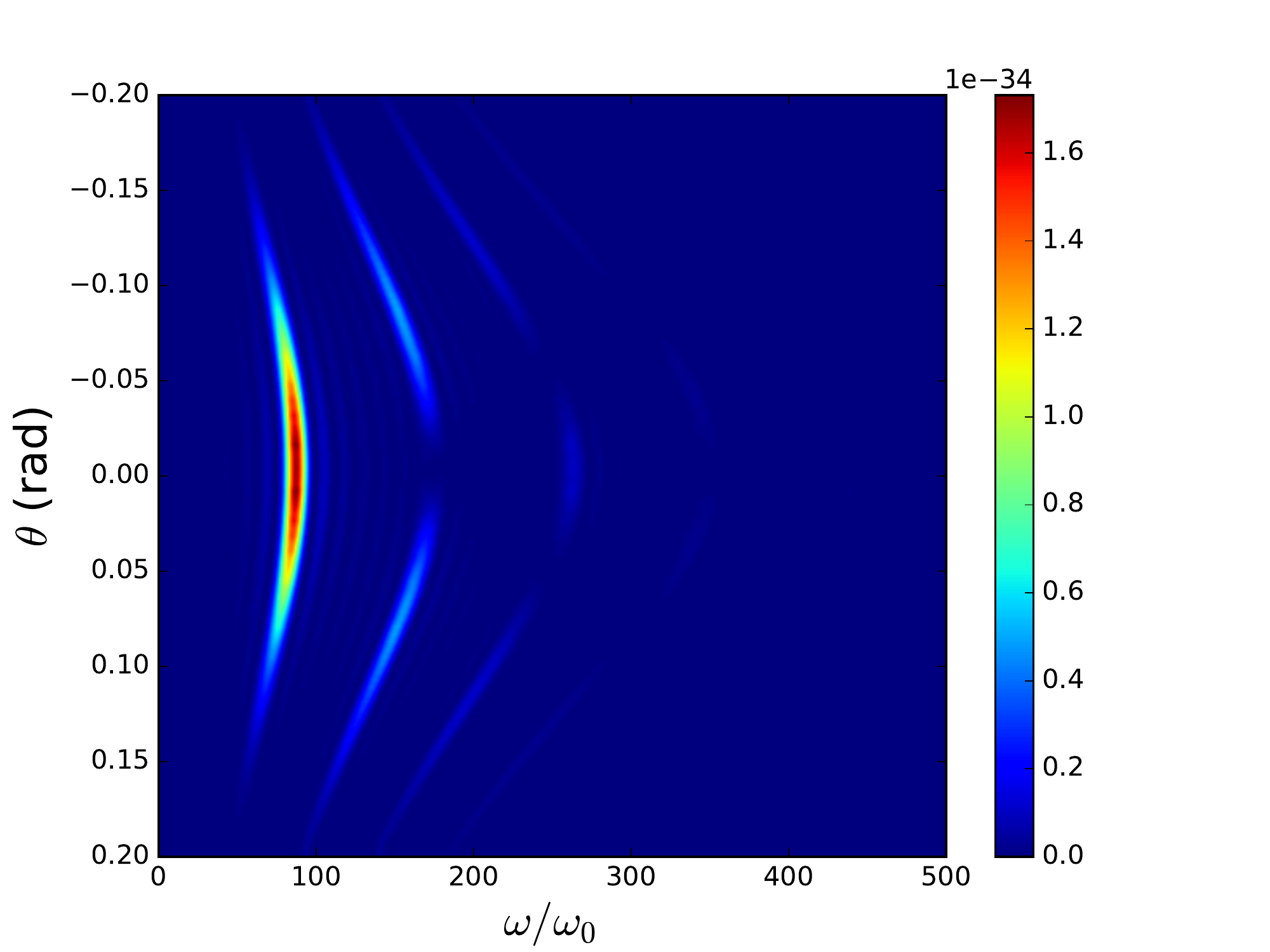}
    \caption{time-domain method}
\end{subfigure}
\begin{subfigure}[b]{0.33\linewidth}%{0.3415\linewidth} this one had been larger
        \includegraphics[width=\textwidth]{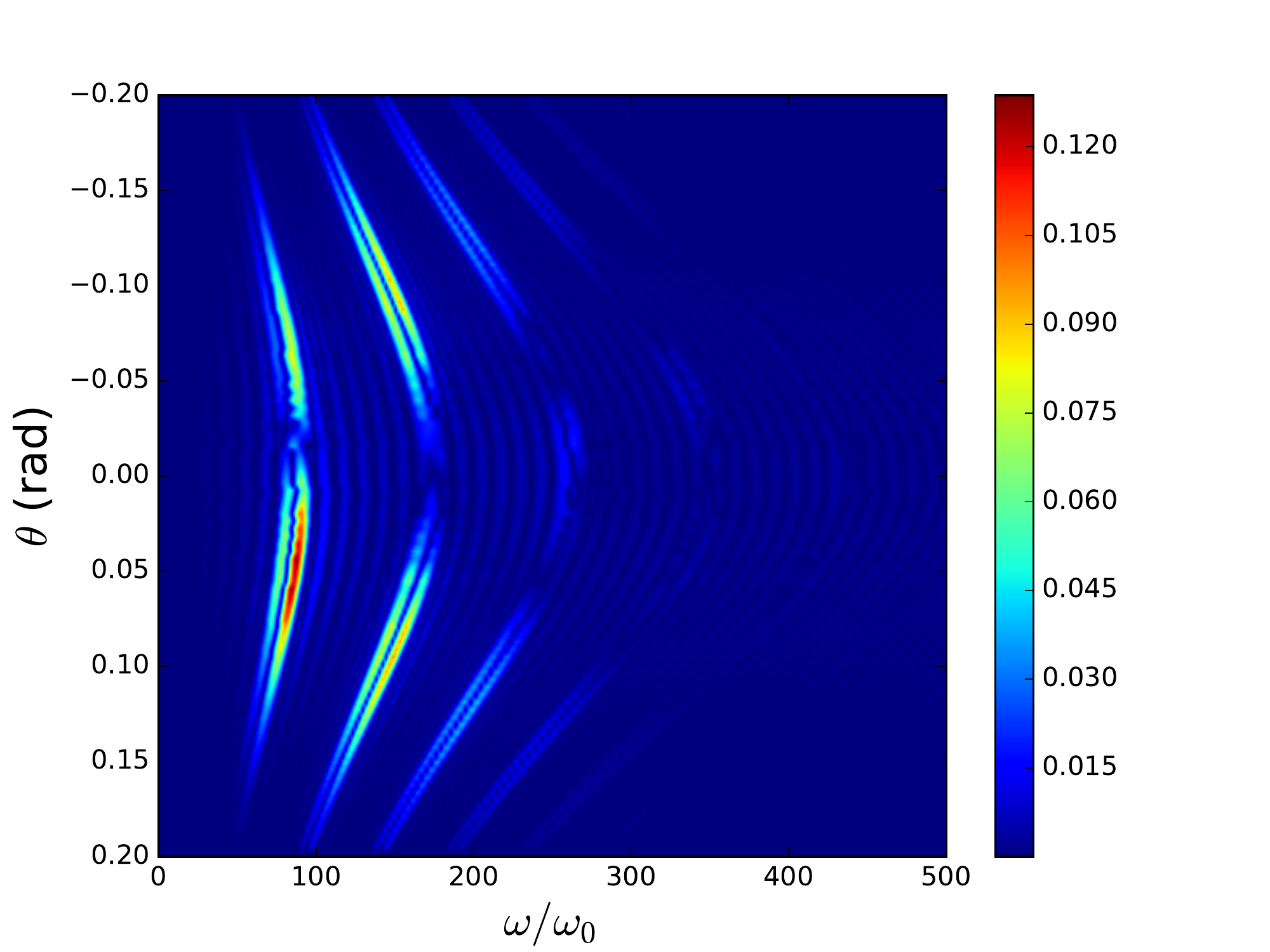}
    \caption{relative error}
\end{subfigure}
\caption{Demonstration of the accuracy of the time-domain method. The radiation spectral-angular distribution is computed by (a) theory and (b) the time-domain method for the configuration stated in section~\ref{sec:comparison:accuracy}. The relative error (c) is computed by Eq.~\eqref{eq:spectra_error}.}
\label{fig:error_full_domain_time}
\end{figure}

\begin{figure}
\begin{subfigure}[b]{0.33\linewidth}
        \includegraphics[width=\textwidth]{fig/accuracy/full_domain/theory.pdf}
    \caption{theory}
\end{subfigure}
\begin{subfigure}[b]{0.33\linewidth}
        \includegraphics[width=\textwidth]{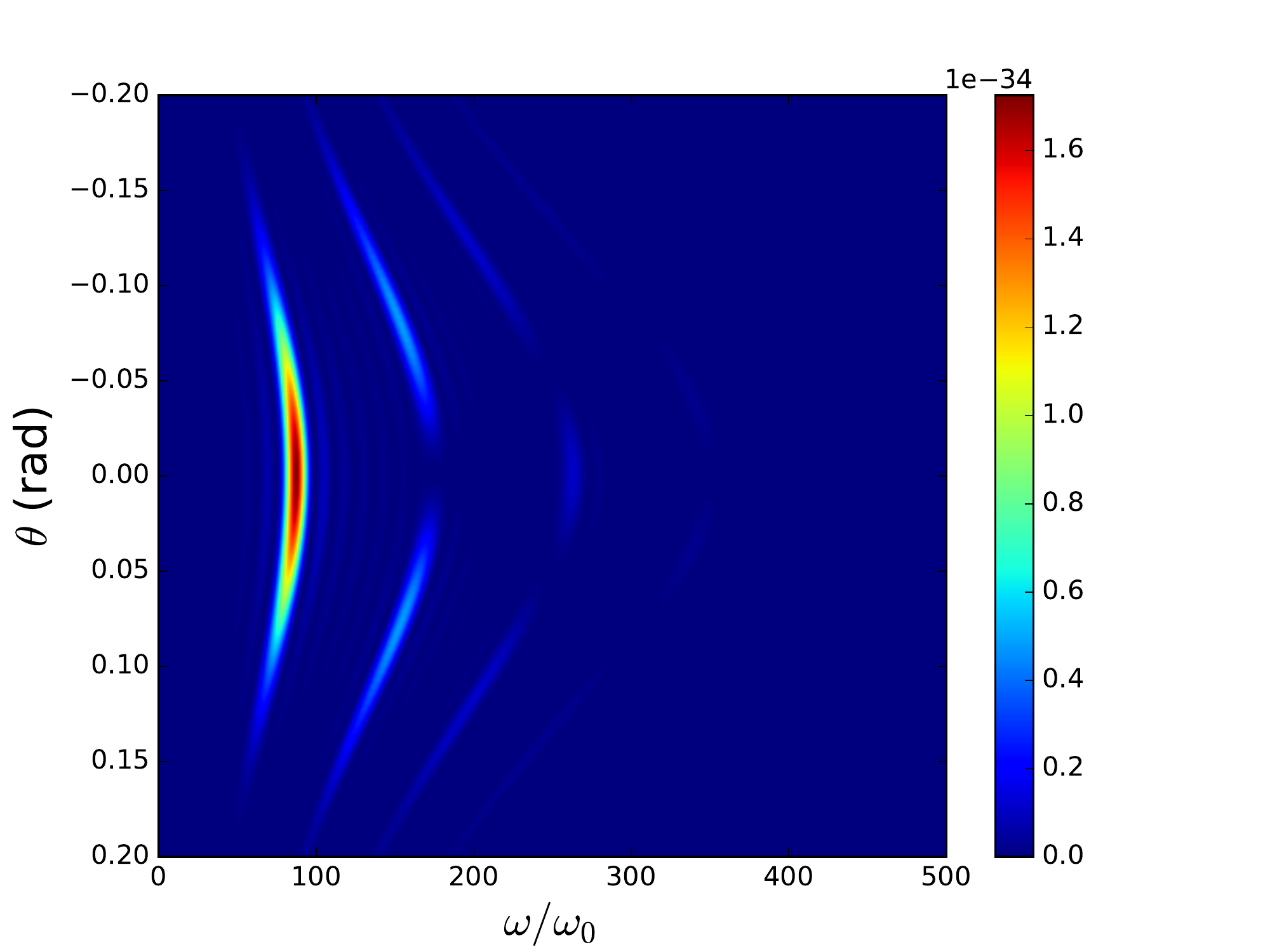}
    \caption{frequency-domain method}
\end{subfigure}
\begin{subfigure}[b]{0.33\linewidth}%{0.3415\linewidth} this one had been larger
        \includegraphics[width=\textwidth]{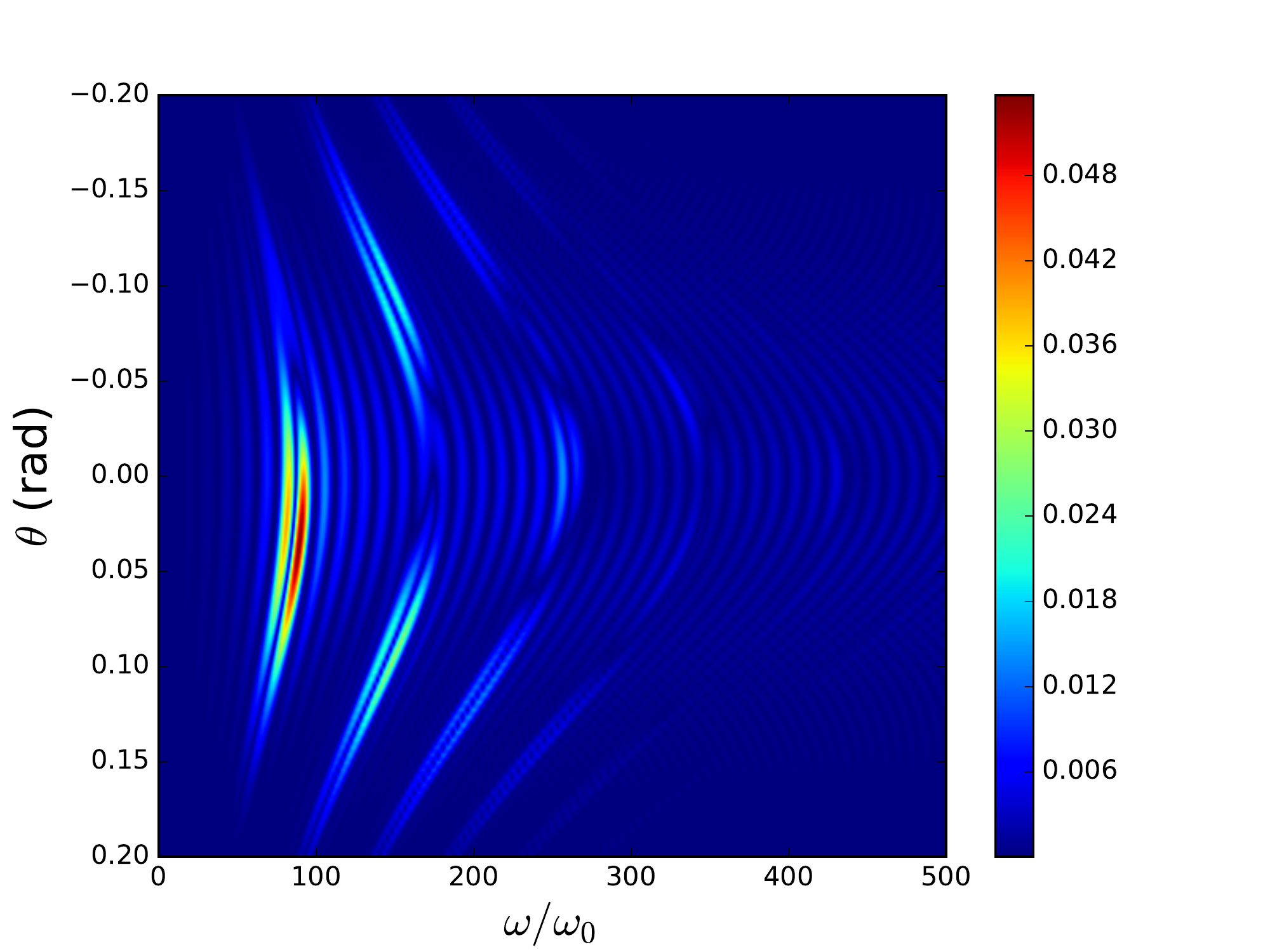}
    \caption{relative error}
\end{subfigure}
\caption{Demonstration of the accuracy of the frequency-domain method. The radiation spectral-angular distribution is computed by (a) theory and (b) the frequency-domain method for the configuration stated in section~\ref{sec:comparison:accuracy}. The relative error (c) is computed by Eq.~\eqref{eq:spectra_error}.}
\label{fig:error_full_domain_freq}
\end{figure}

We perform a single particle simulation in which a particle moves with initial energy $\gamma=5$ in the $+z$ direction and collides with $7$ periods of a counter-propagating sinusoidal wave. 
The radiation is collected in observation directions in the y-z plane (i.\,e., $\theta_x=0$ and $\theta_y = \theta$ in Fig.~\ref{fig:nvector}). 
The radiation spectral-angular distribution and the corresponding errors computed by Eq.~\eqref{eq:spectra_error} are illustrated in Fig.~\ref{fig:error_full_domain_time} for the time-domain method and in Fig.~\ref{fig:error_full_domain_freq} for the frequency-domain method. 

In addition, the maximum and mean values of normalized errors for the radiation spectral-angular distribution with different numbers of frequency points $N_\omega$ and different numbers of observation angles $N_{\theta}$ are shown in Table~\ref{tb:errors_TDM_FDM_Nw} and Table~\ref{tb:errors_TDM_FDM_Nobs}, respectively. 
The relative error over the full spectral-angular distribution for the time-domain method and frequency-domain method has an asymmetric distribution with respect to $\theta=0$. 
This is due to the discretized particle trajectory in the simulation and can be reduced by decreasing the step size for solving the particle trajectory \cite{pausch_how_2014}. From Table~\ref{tb:errors_TDM_FDM_Nobs} and Table~\ref{tb:errors_TDM_FDM_Nw}, we see that both methods achieve an acceptable relative error and that the time-domain method reaches the same level of accuracy as the frequency-domain method by increasing the number of observation points.

% \begin{figure}
% \centering
% \begin{subfigure}[b]{0.45\linewidth}
%         \includegraphics[width=\textwidth]{fig/accuracy/error_max_mean/error_vs_nw.pdf}
%     \caption{}
% \end{subfigure}
% \begin{subfigure}[b]{0.45\linewidth}
%         \includegraphics[width=\textwidth]{fig/accuracy/error_max_mean/error_vs_nobs.pdf}
%     \caption{}
% \end{subfigure}
% \caption{Normalized relative error of time-domain method (diamonds) and frequency-domain method (circles) for (a) different number of frequency points and (b) different number of observation angles.}
% \label{fig:error_max_mean}
% \end{figure}

\begin{table}\renewcommand{\arraystretch}{1.3}
\centering
\begin{tabular}{lccc}\hline\hline
\multicolumn{1}{c}{$N_{\text{obs}}=101$} & $N_{\omega}=5\times 10^3$ & $N_{\omega}=1\times 10^4$ & $N_{\omega}=2\times 10^4$\\\hline
\multicolumn{4}{c}{\textbf{TDM}}\\\hline
$\text{max}(\text{error}(\omega, \theta))$ & $1.287\times 10^{-1}$ & $1.290\times 10^{-1}$ & $1.291\times 10^{-1}$  \\
$\text{mean}(\text{error}(\omega, \theta))$ & $3.520\times 10^{-3}$ & $3.522\times 10^{-3}$ & $3.524\times 10^{-3}$  \\\hline
\multicolumn{4}{c}{\textbf{FDM}} \\\hline
$\text{max}(\text{error}(\omega, \theta))$ & $5.366\times 10^{-2}$ & $5.366\times10^{-2}$ & $5.366 \times 10^{-2}$  \\
$\text{mean}(\text{error}(\omega, \theta))$ & $1.589\times 10^{-3}$ & $1.589\times 10^{-3}$ & $1.580\times 10^{-3}$  \\
\hline
\end{tabular}
\caption{Normalized relative error of the time-domain method (TDM) and the frequency-domain method (FDM) for different numbers of frequency points.}
\label{tb:errors_TDM_FDM_Nw}
\end{table}\renewcommand{\arraystretch}{1}

\begin{table}\renewcommand{\arraystretch}{1.3}
\centering
\begin{tabular}{lccc}\hline\hline
\multicolumn{1}{c}{$N_{\omega}=5\times 10^3$} & $N_{\text{obs}}=101$ & $N_{\text{obs}}=201$ & $N_{\text{obs}}=401$\\\hline
\multicolumn{4}{c}{\textbf{TDM}} \\\hline
$\text{max}(\text{error}(\omega, \theta))$ & $1.291\times 10^{-1}$ & $8.587\times 10^{-2}$ & $6.609\times 10^{-2}$  \\
$\text{mean}(\text{error}(\omega, \theta))$ & $3.524\times 10^{-3}$ & $2.175\times 10^{-3}$ & $1.547\times 10^{-3}$ \\\hline
\multicolumn{4}{c}{\textbf{FDM}}\\\hline
$\text{max}(\text{error}(\omega, \theta))$ & $5.366\times 10^{-2}$ & $5.366\times10^{-2}$ & $5.366 \times 10^{-2}$  \\
$\text{mean}(\text{error}(\omega, \theta))$ & $1.580\times 10^{-3}$ & $1.596\times 10^{-3}$ & $1.599\times 10^{-3}$ \\\hline
\end{tabular}
\caption{Normalized relative error of the time-domain method (TDM) and the frequency-domain method (FDM) for different numbers of observation points.}
\label{tb:errors_TDM_FDM_Nobs}
\end{table}\renewcommand{\arraystretch}{1}

\subsection{Performance}
We will now discuss the computational complexity of the two schemes with respect to the following parameters:
\begin{center}
\begin{tabular}{ll}
	$N_{s}$ & the number of simulation time steps,\\
	$N_{p}$ & the number of particles,\\
	$N_{\text{obs}}$& the number of observers,\\
	$L_t$ & the operation count to compute the advanced time $t_a$, see Eq.~\eqref{eq:far_adv},\\
	$L_f$ & the operation count to evaluate the radiation field, see Eq.~\eqref{eq:far_efield},\\
	$L_I$ & the operation count to interpolate the particle field,\\
	$N_{T_u}$ & the number of grid points of the uniform time grid,\\
	$N_{\omega}$ & the number of grid points of the uniform frequency grid,\\
	$L_{\omega}$ & the operation count to evaluate a particle's contribution in the frequency-domain method, see Eq.~\eqref{eq:freq:far_field_freq}.
\end{tabular}
\end{center}
The total operation count for the time-domain method is given by
\begin{equation}\label{eq:time:op_count_time}
  \text{OP}_{t} = \text{const}\cdot N_{s}\cdot N_p\cdot N_{\text{obs}}\cdot (L_t+L_f+L_I) + \text{const}\cdot N_{\text{obs}}\cdot N_{T_u}\cdot \log N_{T_u},
\end{equation}
whereas for the frequency-domain method we have
\begin{equation}\label{eq:time:op_count_freq}
    \text{OP}_{\omega} =
	\text{const}\cdot N_{s}\cdot N_{p}\cdot N_{\text{obs}} \cdot N_{\omega}\cdot L_{\omega}.	
\end{equation}

\begin{figure}[H]
    \centering
    \begin{subfigure}[b]{0.5\linewidth}
        \includegraphics[width=\textwidth]{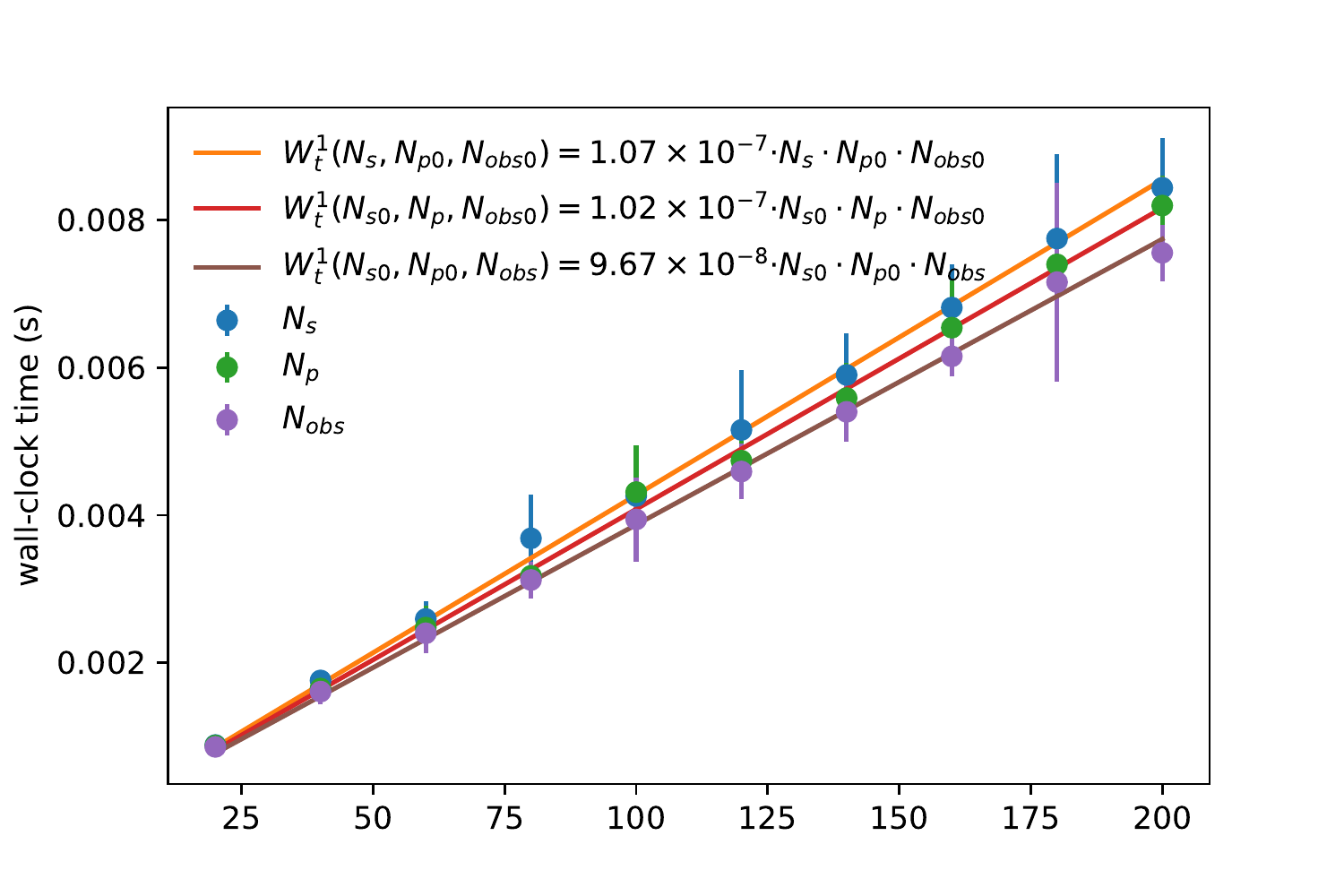}
    \caption{Fitting of $W^1_{t}$}
    \end{subfigure}\\
    \begin{subfigure}[b]{\linewidth}
        \includegraphics[width=\textwidth]{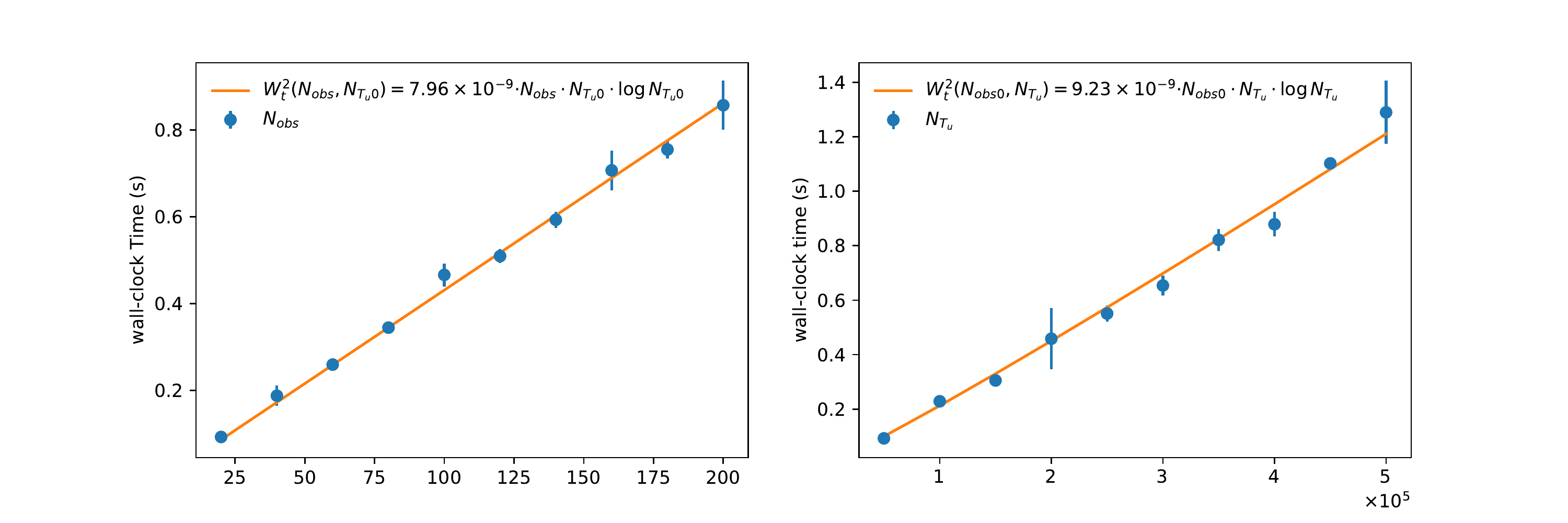}
    \caption{Fitting of $W^2_{t}$}
    \end{subfigure}
    \caption{Fitting of (a) $W^1_{t}$ and (b) $W^2_{t}$ by splitting the multi-variate problem to multiple single-variate fitting problems. Here, $N_{s0}=20$, $N_{p0}=20$, $N_{obs0}=20$ and $N_{T_u0}=5\times 10^4$. Each data point represents the average wall-clock time of $50$ runs. The leading constants $C^{1}_{t}=1.02\times 10^{-7}$ and $C^{2}_{t}=8.61\times 10^{-9}$ are determined as the averages of all respective single-variate fitting constants.}
    \label{fig:wall-clock-fitting-time-domain}
\end{figure}

\begin{figure}[H]
    \centering
    \includegraphics[width=0.5\linewidth]{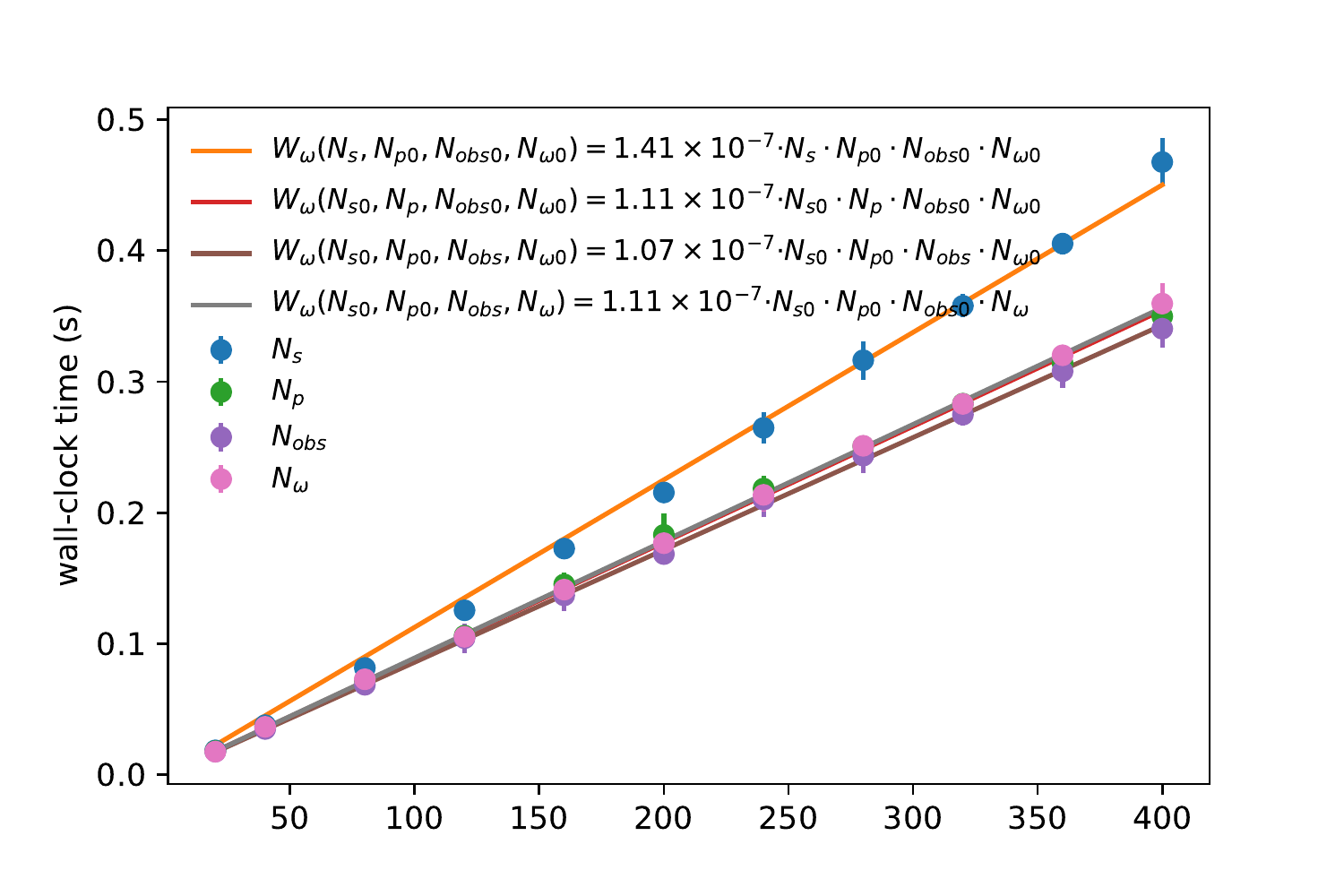}
    \caption{Fitting of $W_{\omega}$ by splitting the multi-variate problem to multiple single-variate fitting problems. $N_{s0}=20$, $N_{p0}=20$, $N_{obs0}=20$ and $N_{\omega 0}=20$. The leading constant $C_{\omega}=1.17\times 10^{-7}$ is determined as the average of all single-variate fitting constants.}
    \label{fig:wall-clock-fitting-frequency-domain}
\end{figure}

The second term in the operation count for the time-domain method stems from an additional post-processing phase in which the radiation data on the uniform time grid is transformed to the frequency-domain by the FFT algorithm in order to obtain the radiation spectra. To easily measure the performance, we define a wall-clock time model for both the time-domain and frequency-domain method. If the memory latency is neglected and the wall-clock time only depends on the operation count of the method, {i.\,e.}, $\text{OP}_{t}$ and $\text{OP}_{\omega}$, the wall-clock time for the time-domain method can be split into the two parts
\begin{align*}
    W^{1}_{t}(N_{s}, N_{p}, N_{\text{obs}}) &= C^{1}_{t}\cdot N_{s}\cdot N_p\cdot N_{\text{obs}}, &
    W^{2}_{t}(N_{\text{obs}}, N_{T_u}) &= C^{2}_{t}\cdot N_{\text{obs}}\cdot N_{T_u}\cdot\log N_{T_u},
\end{align*}
and the wall-clock time for the frequency-domain method is denoted by
\begin{equation*}
    W_{\omega}(N_{s}, N_{p}, N_{\text{obs}}, N_{\omega})= C_{\omega}\cdot N_{s}\cdot N_{p}\cdot N_{\text{obs}}\cdot N_{\omega}.
\end{equation*}
The leading constants translate the operation count into wall-clock time and depend on the implementation and computer architecture (\emph{e.g.}, implementation details, operating system, hardware, and compiler, which are not our primary concerns in this study). Since our analysis only involves the ratio of the constants, the dependency of the computer architecture at which the simulation is performed will most likely be cancelled out.
We can determine the leading constants for both models by the regression of several benchmark runs of both methods (Fig.~\ref{fig:wall-clock-fitting-time-domain} and Fig.~\ref{fig:wall-clock-fitting-frequency-domain}) and obtain
\begin{equation} 
C^{1}_{t}=1.02\times 10^{-7}, \quad C^{2}_{t}=8.61\times 10^{-9}, \quad C_{\omega}=1.17\times 10^{-7}\label{eq:work_constants}
\end{equation}
for our implementation and computer architecture.

The wall-clock time ratio of the time-domain and frequency-domain method is given by
\begin{equation}
    \dfrac{W^{1}_{t} +W^{2}_{t}}{W_{\omega}}=\dfrac{C^{1}_{t}\cdot N_{s}\cdot N_{p} + C^{2}_{t}\cdot N_{T_u}\cdot \log N_{T_u} }{C_{\omega}\cdot N_{s}\cdot N_{p}\cdot N_{\omega}}.
    \label{eq:performance_model_ratio}
\end{equation}
If we set $N_{T_u}=2N_{\omega}$ for the reason of sampling theory, the ratio simplifies to
\begin{equation*}
    \dfrac{W^{1}_{t} +W^{2}_{t}}{W_{\omega}}=
    \dfrac{C^{1}_{t}}{C_{\omega}}\dfrac{1}{N_{\omega}}  + \dfrac{C^{2}_{t}}{C_{\omega}}\dfrac{2}{N_{s}N_{p}}\log 2N_{\omega}.
\end{equation*}
The ratio is greater than one (\emph{i.e.}, the frequency-domain method outperforms the time-domain method) if 
\begin{equation*}
    N_{\omega} > \dfrac{1}{2}\exp\left(\dfrac{1}{2}\dfrac{C_{\omega}}{C^{2}_{t}}\cdot N_{s}\cdot N_{p}\right),
\end{equation*}
which is typically not satisfied with the parameters from practical scenarios. For a simulation of a 1pC particle bunch colliding with a \SI{1}{\micro\metre} laser pulse with $1ps$ duration, the required simulation parameters are $N_{s}=6000$ ($0.167$ femtosecond per time step) and $N_{p}=6.25\times 10^6$. In this case, using the experimentally fitted constants in \eqref{eq:work_constants}, the criterion for the frequency-domain method to outperform the time-domain method is $N_{\omega} > 0.5 \exp(2.5479\cdot 10^{11}) = {\cal O}\left(10^{10^{11}}\right)$.

In the previous discussion, we assumed that the number of time points for the time-domain method is twice the number of frequency points for the frequency-domain method which originated from the assumption that the frequency resolutions for both methods are the same. 
However, this condition is not necessary in practice. 
For the time-domain method, the upper limit of $\Delta \omega$ is restrictively determined (due to the sampling theorem) by the total radiation pulse duration. 
It could be possible that the upper limit of $\Delta \omega$ is far less than the actual resolution that we need to study a problem. 
For the frequency-domain method, on the other hand, there is no such limitation. 
The resolution $\Delta \omega$ can be chosen arbitrarily. 
The bandwidth $\overline{\omega}$ of the maximum frequency which refers to the first harmonic frequency of ICS is \cite{esarey_nonlinear_1993} 
\begin{equation*}
    \overline{\omega} = \dfrac{\lambda_0}{c_0 T_{\text{laser}}}\omega_{\text{max}}.
\end{equation*}
Thus, when applying the frequency-domain method,
\begin{equation}
    \Delta \omega = \dfrac{1}{N_{\overline{\omega}}}
    \dfrac{\lambda_0}{c_0 T_{\text{laser}}}\omega_{\text{max}}
    \label{eq:frequency_resolution_for_ics}
    \quad\text{and}\quad 
    N_{\omega} = N_{\overline{\omega}}
    \dfrac{c_0 T_{\text{laser}}}{\lambda_0},
\end{equation}
where $N_{\overline{\omega}}$ is the number of grid points needed for $\overline{\omega}$. To find a condition under which the time-domain method is slower than the frequency-domain method, Eq.~\eqref{eq:performance_model_ratio} can be expressed as
\begin{equation*}
    C^{2}_{t}\cdot N_{T_u}\cdot\log N_{T_u} > (C_{\omega}\cdot N_{\omega}-C^{1}_{t})N_{s}\cdot N_{p}
\end{equation*}
and rewritten by
\begin{equation}
    N_{T_u}\cdot\log N_{T_u} > \dfrac{C_{\omega}}{C^{2}_{t}}N_{\omega}\cdot N_{s}\cdot N_{p}.
    \label{eq:performance_model_favorable_condition}
\end{equation}
Here, $C^{1}_{t}$ should be of the same order of magnitude as $C_{\omega}$ and is negligible compared to $C_{\omega}N_{\omega}$. From Eq.~\eqref{eq:performance_model_favorable_condition}, we can have two immediate conclusions:
\begin{compactenum}
    \item The time-domain method is favorable when the laser pulse duration or the charge of the particle beam is large.
    \item The frequency-domain method is better when a particle beam with high energy or long bunch length is considered.
\end{compactenum}
In addition, the factor $C_{\omega}/C^2_{t}$ in Eq.~\eqref{eq:performance_model_favorable_condition} can also have significant contribution. The value of $C_{\omega}$ depends not only on the performance of hardware but also on the algorithm for solving the particle trajectory. If a more sophisticated algorithm is utilized (\emph{e.g.}, particle-in-cell method \cite{haugbolle_photon-plasma_2013, pausch_computing_2014}), the time-domain method may become more favorable. In Table~\ref{tb:performance_estimation_cases}, we demonstrate the ratio of $N_{T_u}\cdot\log N_{T_u}$ and $C_{\omega}/C^{2}_{t}\cdot N_{\omega}\cdot N_{s}\cdot N_{p}$ for different experimental projects. We can observe that the time-domain method is still faster than the frequency-domain method even if we choose $N_{\omega}$ so that $\Delta \omega$ fulfills Eq.~\eqref{eq:frequency_resolution_for_ics}.

\begin{figure}[H]
    \centering
    \includegraphics[width=0.5\linewidth]{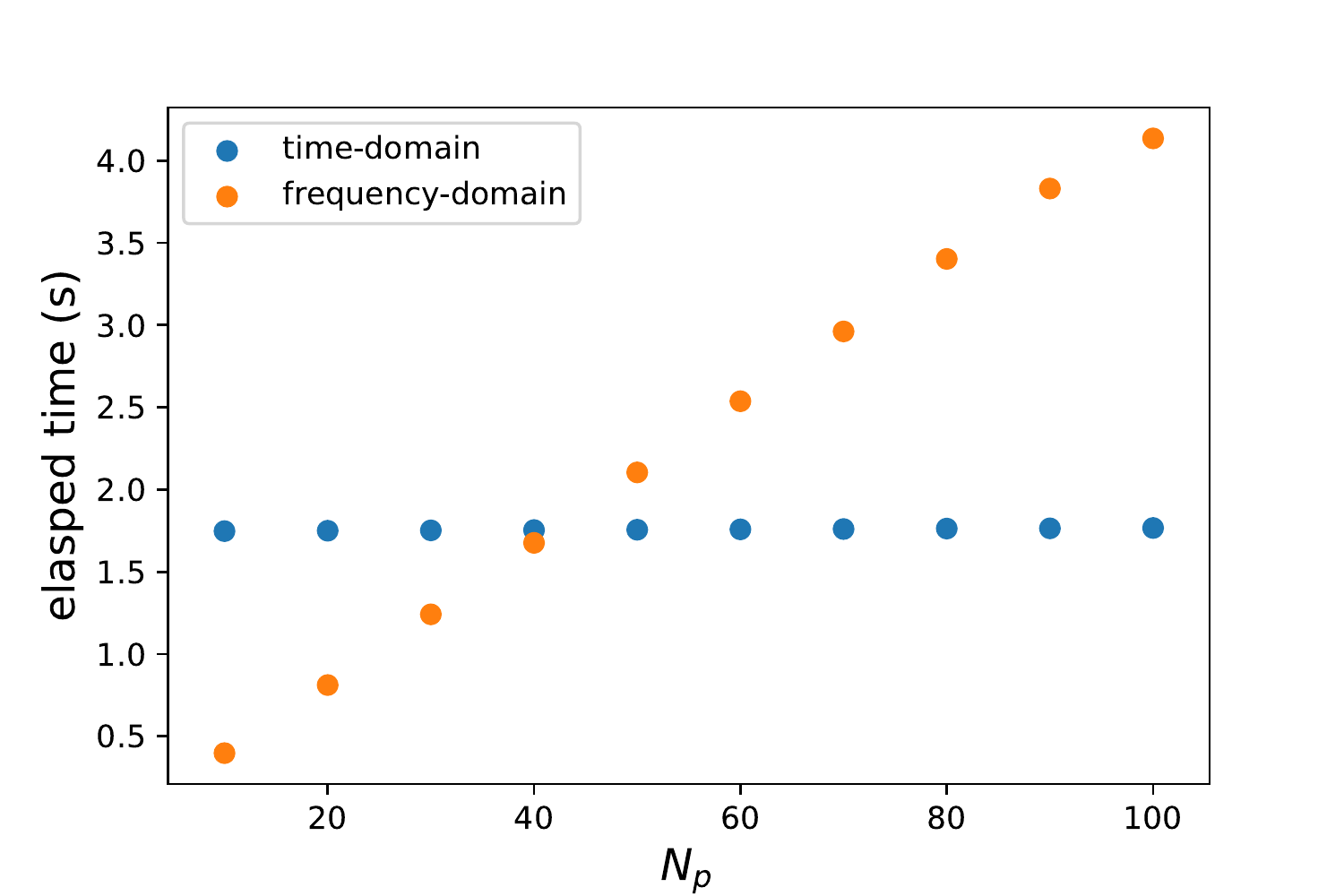}
    \caption{Elapsed times for the simulation ($N_{s}=100$) with different numbers of particles ($N_{p}$) by the time-domain method ($N_{Tu}=1.28\times 10^6$) and the frequency-domain method ($N_{\omega}=100$). Each data point is the average elapsed time from 50 simulation runs. In this case, the frequency-domain method outperforms the time-domain method when $N_{p}\leq 40$. }
    \label{fig:performance_transition}
\end{figure}

Although the TDM outperforms the FDM for large particle numbers typically used in ICS sources, it is interesting to study the transition where the TDM outperforms the FDM for low particle numbers since ICS is also used in other applications. To numerically demonstrate the situation where FDM outperforms the TDM, we consider a bunch of particles uniformly distributed along the longitudinal direction with length of $\SI{100}{\micro\metre}$. This particle bunch interacts with 5 periods of a \SI{1}{\micro\metre} sinusoidal wave. We measure the elapsed time for simulation with different numbers of particles by both TDM and FDM. The result is demonstrated in Fig.~\ref{fig:performance_transition}, the FDM wins when $N_{p}\leq 40$ and the TDM wins when $N_{p} > 40$.

Although the performance model is verified by the execution times in serial, our conclusions can be extended to the scenario with parallelization (i.\,e. with detector parallelization) in which the total operation counts for the time-domain and frequency-domain methods can be written as $\text{OP}_t/P$ and $\text{OP}_{\omega}/P$, resp..

\begin{table}\renewcommand{\arraystretch}{1.1}
\centering
\begin{tabular}{cccccccc}\hline\hline
AXSIS & ODU CLS & ASU CXFEL & ASU CXLS & XFI & MuLCS & ThomX\\
\hline
$4.75\times 10^{-7}$ & $6.09\times 10^{-9}$ & $3.77\times 10^{-10}$ & $1.69\times 10^{-10}$ & $1.89\times 10^{-10}$ &
$3.91\times 10^{-11}$ &
$1.79\times 10^{-10}$\\
\hline
\end{tabular}
\caption{The ratio of $N_{T_u}\cdot\log N_{T_u}$ and $C_{\omega}/C^{2}_{t}\cdot N_{\omega}\cdot N_{s}\cdot N_{p}$ for different experimental projects is computed. Here, the electron beam transverse size $l_x$ and $l_y$ are not considered in the evaluation of $N_{Tu}$ as their contribution is minor for a high energy electron beam. The reference of each experimental project can be found in Table~\ref{tb:Ntu_Ns_ratio_for_ics_experiments}.}
\label{tb:performance_estimation_cases}
\end{table}\renewcommand{\arraystretch}{1}

%\section{Outlook}
In this study, the space charge force between charged particles has not been included when solving the dynamics of charged particles. However, it might be significant for some light sources from particle beams with relatively low $\gamma$ (\emph{e.g.} AXSIS \cite{kartner_axsis_2016}), and will be the subject of future work.
%\done{JZ: I guess it is a good idea to include some sketch of the inclusion of particle interactions in ICS as the next step?}

\section*{Acknowledgement}
The authors acknowledge the support by DASHH (Data Science in Hamburg - HELMHOLTZ Graduate School for the Structure of Matter) with the Grant-No. HIDSS-0002. This research was supported in part through the Maxwell computational resources operated at Deutsches Elektronen-Synchrotron (DESY), Hamburg, Germany.


 \bibliographystyle{elsarticle-num} 
 \bibliography{reference}

\begin{thebibliography}{10}
\expandafter\ifx\csname url\endcsname\relax
  \def\url#1{\texttt{#1}}\fi
\expandafter\ifx\csname urlprefix\endcsname\relax\def\urlprefix{URL }\fi
\expandafter\ifx\csname href\endcsname\relax
  \def\href#1#2{#2} \def\path#1{#1}\fi

\bibitem{bostedt_linac_2016}
C.~Bostedt, S.~Boutet, D.~M. Fritz, Z.~Huang, H.~J. Lee, H.~T. Lemke,
  A.~Robert, W.~F. Schlotter, J.~J. Turner, G.~J. Williams,
  \href{https://link.aps.org/doi/10.1103/RevModPhys.88.015007}{Linac {Coherent}
  {Light} {Source}: {The} first five years}, Rev. Mod. Phys. 88~(1) (2016)
  015007.
\newblock \href {https://doi.org/10.1103/RevModPhys.88.015007}
  {\path{doi:10.1103/RevModPhys.88.015007}}.
\newline\urlprefix\url{https://link.aps.org/doi/10.1103/RevModPhys.88.015007}

\bibitem{pellegrini_development_2020}
C.~Pellegrini, \href{http://www.nature.com/articles/s42254-020-0197-1}{The
  development of {XFELs}}, Nat Rev Phys 2~(7) (2020) 330--331.
\newblock \href {https://doi.org/10.1038/s42254-020-0197-1}
  {\path{doi:10.1038/s42254-020-0197-1}}.
\newline\urlprefix\url{http://www.nature.com/articles/s42254-020-0197-1}

\bibitem{krafft_compton_2010}
G.~A. Krafft, G.~Priebe,
  \href{https://www.worldscientific.com/doi/abs/10.1142/S1793626810000440}{Compton
  {Sources} of {Electromagnetic} {Radiation}}, Rev. Accl. Sci. Tech. 03~(01)
  (2010) 147--163.
\newblock \href {https://doi.org/10.1142/S1793626810000440}
  {\path{doi:10.1142/S1793626810000440}}.
\newline\urlprefix\url{https://www.worldscientific.com/doi/abs/10.1142/S1793626810000440}

\bibitem{ta_phuoc_all-optical_2012}
K.~Ta~Phuoc, S.~Corde, C.~Thaury, V.~Malka, A.~Tafzi, J.~P. Goddet, R.~C. Shah,
  S.~Sebban, A.~Rousse,
  \href{http://www.nature.com/articles/nphoton.2012.82}{All-optical {Compton}
  gamma-ray source}, Nature Photon 6~(5) (2012) 308--311.
\newblock \href {https://doi.org/10.1038/nphoton.2012.82}
  {\path{doi:10.1038/nphoton.2012.82}}.
\newline\urlprefix\url{http://www.nature.com/articles/nphoton.2012.82}

\bibitem{bilderback_review_2005}
D.~H. Bilderback, P.~Elleaume, E.~Weckert,
  \href{https://iopscience.iop.org/article/10.1088/0953-4075/38/9/022}{Review
  of third and next generation synchrotron light sources}, J. Phys. B: At. Mol.
  Opt. Phys. 38~(9) (2005) S773--S797.
\newblock \href {https://doi.org/10.1088/0953-4075/38/9/022}
  {\path{doi:10.1088/0953-4075/38/9/022}}.
\newline\urlprefix\url{https://iopscience.iop.org/article/10.1088/0953-4075/38/9/022}

\bibitem{graves_compact_2014}
W.~Graves, J.~Bessuille, P.~Brown, S.~Carbajo, V.~Dolgashev, K.-H. Hong,
  E.~Ihloff, B.~Khaykovich, H.~Lin, K.~Murari, E.~Nanni, G.~Resta, S.~Tantawi,
  L.~Zapata, F.~Kärtner, D.~Moncton,
  \href{https://link.aps.org/doi/10.1103/PhysRevSTAB.17.120701}{Compact x-ray
  source based on burst-mode inverse {Compton} scattering at 100 {kHz}}, Phys.
  Rev. ST Accel. Beams 17~(12) (2014) 120701.
\newblock \href {https://doi.org/10.1103/PhysRevSTAB.17.120701}
  {\path{doi:10.1103/PhysRevSTAB.17.120701}}.
\newline\urlprefix\url{https://link.aps.org/doi/10.1103/PhysRevSTAB.17.120701}

\bibitem{kartner_axsis_2016}
F.~Kärtner, F.~Ahr, A.-L. Calendron, H.~Çankaya, S.~Carbajo, G.~Chang,
  G.~Cirmi, K.~Dörner, U.~Dorda, A.~Fallahi, A.~Hartin, M.~Hemmer, R.~Hobbs,
  Y.~Hua, W.~Huang, R.~Letrun, N.~Matlis, V.~Mazalova, O.~Mücke, E.~Nanni,
  W.~Putnam, K.~Ravi, F.~Reichert, I.~Sarrou, X.~Wu, A.~Yahaghi, H.~Ye,
  L.~Zapata, D.~Zhang, C.~Zhou, R.~Miller, K.~Berggren, H.~Graafsma, A.~Meents,
  R.~Assmann, H.~Chapman, P.~Fromme,
  \href{https://linkinghub.elsevier.com/retrieve/pii/S0168900216002564}{{AXSIS}:
  {Exploring} the frontiers in attosecond {X}-ray science, imaging and
  spectroscopy}, Nuclear Instruments and Methods in Physics Research Section A:
  Accelerators, Spectrometers, Detectors and Associated Equipment 829 (2016)
  24--29.
\newblock \href {https://doi.org/10.1016/j.nima.2016.02.080}
  {\path{doi:10.1016/j.nima.2016.02.080}}.
\newline\urlprefix\url{https://linkinghub.elsevier.com/retrieve/pii/S0168900216002564}

\bibitem{esarey_nonlinear_1993}
E.~Esarey, S.~K. Ride, P.~Sprangle,
  \href{https://link.aps.org/doi/10.1103/PhysRevE.48.3003}{Nonlinear {Thomson}
  scattering of intense laser pulses from beams and plasmas}, Phys. Rev. E
  48~(4) (1993) 3003--3021.
\newblock \href {https://doi.org/10.1103/PhysRevE.48.3003}
  {\path{doi:10.1103/PhysRevE.48.3003}}.
\newline\urlprefix\url{https://link.aps.org/doi/10.1103/PhysRevE.48.3003}

\bibitem{deitrick_high_2017}
K.~Deitrick, J.~Delayen, G.~Krafft,
  \href{http://jacow.org/ipac2017/papers/mopva036.pdf}{{H}igh {A}verage
  {B}rilliance {C}ompact {I}nverse {C}ompton {L}ight {S}ource}, in: Proc. of
  International Particle Accelerator Conference (IPAC'17), Copenhagen, Denmark,
  from May 14 – 19, 2017, no.~8 in International Particle Accelerator
  Conference, JACoW, Geneva, Switzerland, 2017, pp. 932--935,
  https://doi.org/10.18429/JACoW-IPAC2017-MOPVA036.
\newblock \href
  {https://doi.org/https://doi.org/10.18429/JACoW-IPAC2017-MOPVA036}
  {\path{doi:https://doi.org/10.18429/JACoW-IPAC2017-MOPVA036}}.
\newline\urlprefix\url{http://jacow.org/ipac2017/papers/mopva036.pdf}

\bibitem{deitrick_high-brilliance_2018-1}
K.~Deitrick, G.~Krafft, B.~Terzi{\'c}, J.~Delayen, High-brilliance, high-flux
  compact inverse compton light source, Physical Review Accelerators and Beams
  21~(8) (2018) 080703.

\bibitem{graves_asu_2018}
W.~Graves, J.~Chen, P.~Fromme, M.~Holl, K.-H. Hong, R.~Kirian,
  C.~Limborg-Deprey, L.~Malin, D.~Moncton, E.~Nanni, K.~Schmidt, J.~Spence,
  M.~Underhill, U.~Weierstall, N.~Zatsepin, C.~Zhang,
  \href{http://jacow.org/fel2017/doi/JACoW-FEL2017-TUB03.html}{{ASU} {Compact}
  {XFEL}}, Proceedings of the 38th Int. Free Electron Laser Conf. FEL2017
  (2018) 4 pages, 0.925 MB, artwork Size: 4 pages, 0.925 MB ISBN: 9783954501793
  Medium: PDF Publisher: JACoW, Geneva, Switzerland.
\newblock \href {https://doi.org/10.18429/JACOW-FEL2017-TUB03}
  {\path{doi:10.18429/JACOW-FEL2017-TUB03}}.
\newline\urlprefix\url{http://jacow.org/fel2017/doi/JACoW-FEL2017-TUB03.html}

\bibitem{brummer_design_2020}
T.~Brümmer, A.~Debus, R.~Pausch, J.~Osterhoff, F.~Grüner,
  \href{https://link.aps.org/doi/10.1103/PhysRevAccelBeams.23.031601}{Design
  study for a compact laser-driven source for medical x-ray fluorescence
  imaging}, Phys. Rev. Accel. Beams 23~(3) (2020) 031601.
\newblock \href {https://doi.org/10.1103/PhysRevAccelBeams.23.031601}
  {\path{doi:10.1103/PhysRevAccelBeams.23.031601}}.
\newline\urlprefix\url{https://link.aps.org/doi/10.1103/PhysRevAccelBeams.23.031601}

\bibitem{gunther_versatile_2020}
B.~Günther, R.~Gradl, C.~Jud, E.~Eggl, J.~Huang, S.~Kulpe, K.~Achterhold,
  B.~Gleich, M.~Dierolf, F.~Pfeiffer,
  \href{https://scripts.iucr.org/cgi-bin/paper?S1600577520008309}{The versatile
  {X}-ray beamline of the {Munich} {Compact} {Light} {Source}: design,
  instrumentation and applications}, J Synchrotron Rad 27~(5) (2020)
  1395--1414, muLCS (ring).
\newblock \href {https://doi.org/10.1107/S1600577520008309}
  {\path{doi:10.1107/S1600577520008309}}.
\newline\urlprefix\url{https://scripts.iucr.org/cgi-bin/paper?S1600577520008309}

\bibitem{variola_thomx_nodate}
A.~Variola, J.~Haissinski, A.~Loulergue, F.~Zomer, e.~),
  \href{http://hal.in2p3.fr/in2p3-00971281}{{ThomX Technical Design Report}},
  Tech. rep., ThomX - Dept. Accelerateurs (2014).
\newline\urlprefix\url{http://hal.in2p3.fr/in2p3-00971281}

\bibitem{dupraz_thomx_2020}
K.~Dupraz, M.~Alkadi, M.~Alves, L.~Amoudry, D.~Auguste, J.-L. Babigeon,
  M.~Baltazar, A.~Benoit, J.~Bonis, J.~Bonenfant, C.~Bruni, K.~Cassou, J.-N.
  Cayla, T.~Chabaud, I.~Chaikovska, S.~Chance, V.~Chaumat, R.~Chiche,
  A.~Cobessi, P.~Cornebise, O.~Dalifard, N.~Delerue, R.~Dorkel, D.~Douillet,
  J.-P. Dugal, N.~El~Kamchi, M.~El~Khaldi, E.~Ergenlik, P.~Favier,
  M.~Fernandez, A.~Gamelin, J.-F. Garaut, L.~Garolfi, P.~Gauron, F.~Gauthier,
  A.~Gonnin, D.~Grasset, E.~Guerard, H.~Guler, J.~Haissinski, E.~Herry,
  G.~Iaquaniello, M.~Jacquet, E.~Jules, V.~Kubytskyi, M.~Langlet,
  T.~Le~Barillec, J.-F. Ledu, D.~Leguidec, B.~Leluan, P.~Lepercq,
  F.~Letellier-Cohen, R.~Marie, J.-C. Marrucho, A.~Martens, C.~Mageur,
  G.~Mercadier, B.~Mercier, E.~Mistretta, H.~Monard, A.~Moutardier, O.~Neveu,
  D.~Nutarelli, M.~Omeich, Y.~Peinaud, Y.~Petrilli, M.~Pichet, E.~Plaige,
  C.~Prévost, P.~Rudnicky, V.~Soskov, M.~Taurigna-Quéré, S.~Trochet,
  C.~Vallerand, O.~Vitez, F.~Wicek, S.~Wurth, F.~Zomer, P.~Alexandre, R.~Ben
  El~Fekih, P.~Berteaud, F.~Bouvet, R.~Cuoq, A.~Diaz, Y.~Dietrich, M.~Diop,
  D.~Pedeau, E.~Dupuy, F.~Marteau, F.~Bouvet, A.~Gamelin, D.~Helder, N.~Hubert,
  J.~Veteran, M.~Labat, A.~Lestrade, A.~Letrésor, R.~Lopes, A.~Loulergue,
  M.~Louvet, M.~Louvet, P.~Marchand, M.~El~Ajjouri, D.~Muller, A.~Nadji,
  L.~Nadolski, R.~Nagaoka, S.~Petit, J.-P. Pollina, F.~Ribeiro, M.~Ros,
  J.~Salvia, S.~Bobault, M.~Sebdaoui, R.~Sreedharan, Y.~Bouanai, J.-L.
  Hazemann, J.-L. Hodeau, E.~Roy, P.~Jeantet, J.~Lacipière, P.~Robert, J.-M.
  Horodynski, H.~Bzyl, C.~Chapelle, M.~Biagini, P.~Walter, A.~Bravin,
  W.~Del~Net, E.~Lahéra, O.~Proux, H.~Elleaume, E.~Cormier,
  \href{https://linkinghub.elsevier.com/retrieve/pii/S2666032620300387}{The
  {ThomX} {ICS} source}, Physics Open 5 (2020) 100051, thomx (Ring).
\newblock \href {https://doi.org/10.1016/j.physo.2020.100051}
  {\path{doi:10.1016/j.physo.2020.100051}}.
\newline\urlprefix\url{https://linkinghub.elsevier.com/retrieve/pii/S2666032620300387}

\bibitem{thomas_algorithm_2010}
A.~G.~R. Thomas,
  \href{https://link.aps.org/doi/10.1103/PhysRevSTAB.13.020702}{Algorithm for
  calculating spectral intensity due to charged particles in arbitrary motion},
  Phys. Rev. ST Accel. Beams 13~(2) (2010) 020702.
\newblock \href {https://doi.org/10.1103/PhysRevSTAB.13.020702}
  {\path{doi:10.1103/PhysRevSTAB.13.020702}}.
\newline\urlprefix\url{https://link.aps.org/doi/10.1103/PhysRevSTAB.13.020702}

\bibitem{frederiksen_radiation_2010}
J.~T. Frederiksen, T.~Haugbølle, M.~V. Medvedev, {\AA}.~Nordlund,
  \href{https://iopscience.iop.org/article/10.1088/2041-8205/722/1/L114}{{RADIATION}
  {SPECTRAL} {SYNTHESIS} {OF} {RELATIVISTIC} {FILAMENTATION}}, ApJ 722~(1)
  (2010) L114--L119.
\newblock \href {https://doi.org/10.1088/2041-8205/722/1/L114}
  {\path{doi:10.1088/2041-8205/722/1/L114}}.
\newline\urlprefix\url{https://iopscience.iop.org/article/10.1088/2041-8205/722/1/L114}

\bibitem{haugbolle_photon-plasma_2013}
T.~Haugbølle, J.~T. Frederiksen, {\AA}.~Nordlund,
  \href{http://aip.scitation.org/doi/10.1063/1.4811384}{{PHOTON}-{PLASMA}: {A}
  modern high-order particle-in-cell code}, Physics of Plasmas 20~(6) (2013)
  062904.
\newblock \href {https://doi.org/10.1063/1.4811384}
  {\path{doi:10.1063/1.4811384}}.
\newline\urlprefix\url{http://aip.scitation.org/doi/10.1063/1.4811384}

\bibitem{pausch_computing_2014}
R.~Pausch, H.~Burau, M.~Bussmann, J.~Couperus, T.~Cowan, A.~Debus, A.~Huebl,
  A.~Irman, A.~Köhler, U.~Schramm, K.~Steiniger, R.~Widera,
  \href{http://jacow.org/IPAC2014/doi/JACoW-IPAC2014-MOPRI069.html}{Computing
  {Angularly}-resolved {Far} {Field} {Emission} {Spectra} in {Particle}-in-cell
  {Codes} using {GPUs}}, Proceedings of the 5th Int. Particle Accelerator Conf.
  IPAC2014 (2014) 4 pages, 0.581 MB, artwork Size: 4 pages, 0.581 MB ISBN:
  9783954501328 Medium: PDF Publisher: JACoW, Geneva, Switzerland.
\newblock \href {https://doi.org/10.18429/JACOW-IPAC2014-MOPRI069}
  {\path{doi:10.18429/JACOW-IPAC2014-MOPRI069}}.
\newline\urlprefix\url{http://jacow.org/IPAC2014/doi/JACoW-IPAC2014-MOPRI069.html}

\bibitem{pausch_how_2014}
R.~Pausch, A.~Debus, R.~Widera, K.~Steiniger, A.~Huebl, H.~Burau, M.~Bussmann,
  U.~Schramm,
  \href{https://linkinghub.elsevier.com/retrieve/pii/S0168900213014642}{How to
  test and verify radiation diagnostics simulations within particle-in-cell
  frameworks}, Nuclear Instruments and Methods in Physics Research Section A:
  Accelerators, Spectrometers, Detectors and Associated Equipment 740 (2014)
  250--256.
\newblock \href {https://doi.org/10.1016/j.nima.2013.10.073}
  {\path{doi:10.1016/j.nima.2013.10.073}}.
\newline\urlprefix\url{https://linkinghub.elsevier.com/retrieve/pii/S0168900213014642}

\bibitem{pausch_electromagnetic_2012}
R.~Pausch, \href{https://doi.org/10.5281/zenodo.843510}{Electromagnetic
  {Radiation} from {Relativistic} {Electrons} as {Characteristic} {Signature}
  of their {Dynamics}}, {PhD} {Thesis}, Technische Universität Dresden (Dec.
  2012).
\newblock \href {https://doi.org/10.5281/zenodo.843510}
  {\path{doi:10.5281/zenodo.843510}}.
\newline\urlprefix\url{https://doi.org/10.5281/zenodo.843510}

\bibitem{sell_attosecond_2014}
A.~Sell, F.~X. Kärtner,
  \href{https://iopscience.iop.org/article/10.1088/0953-4075/47/1/015601}{Attosecond
  electron bunches accelerated and compressed by radially polarized laser
  pulses and soft-x-ray pulses from optical undulators}, J. Phys. B: At. Mol.
  Opt. Phys. 47~(1) (2014) 015601.
\newblock \href {https://doi.org/10.1088/0953-4075/47/1/015601}
  {\path{doi:10.1088/0953-4075/47/1/015601}}.
\newline\urlprefix\url{https://iopscience.iop.org/article/10.1088/0953-4075/47/1/015601}

\bibitem{jackson_classical_nodate}
D.~Jackson, Classical {Electrodynamics}, Wiley, 1999.

\bibitem{ryne_using_2013}
R.~D. Ryne, Using a {Lienard}-{Wiechert} {Solver} to {Study} {Coherent}
  {Synchrotron} {Radiation} {Effects}, New York (2013) 7.

\bibitem{boris1970}
J.~P. Boris, Relativistic plasma simulation-optimization of a hybrid code, in:
  Proc. Fourth Conf. Num. Sim. Plasmas, Naval Res. Lab, Wash. DC, 1970, pp.
  3--67.

\bibitem{deitrick_inverse_2017}
K.~E. Deitrick, Inverse Compton light source: A compact design proposal, Old
  Dominion University, 2017.

\bibitem{deitrick_high-brilliance_2018}
K.~Deitrick, G.~Krafft, B.~Terzić, J.~Delayen,
  \href{https://link.aps.org/doi/10.1103/PhysRevAccelBeams.21.080703}{High-brilliance,
  high-flux compact inverse {Compton} light source}, Phys. Rev. Accel. Beams
  21~(8) (2018) 080703, old domin (Linac).
\newblock \href {https://doi.org/10.1103/PhysRevAccelBeams.21.080703}
  {\path{doi:10.1103/PhysRevAccelBeams.21.080703}}.
\newline\urlprefix\url{https://link.aps.org/doi/10.1103/PhysRevAccelBeams.21.080703}

\end{thebibliography}





% \end{thebibliography}
\end{document}